\documentclass{aa}  
\usepackage{aas_macros}
\usepackage{graphicx}
\usepackage[export]{adjustbox} 
\usepackage{comment}
\usepackage{subfigure}
\usepackage{txfonts}
\usepackage{xcolor}

\usepackage{hyperref}
\hypersetup{
    colorlinks=true,
    linkcolor=blue,
    citecolor=blue,
    filecolor=magenta,      
    urlcolor=blue,
    pdfauthor={Firinu et al., 2026},
    pdftitle={The Impact of the Magnetised Cosmic Web on UHECR Propagation}
    pdfpagemode={UseOutlines},
    }

\begin{document} 
\title{The Impact of the Magnetised Cosmic Web on Ultra High Energy Cosmic Ray Propagation}

\author{A.~Firinu \inst{1}, F.~Vazza \inst{1,2}, C.~Evoli \inst{3,4}}
\offprints{%
E-mail:}
\institute{Dipartimento di Fisica e Astronomia, Universit\`{a} di Bologna, Via Gobetti 93/2, 40122, Bologna, Italy 
\and Istituto di Radio Astronomia, Isitituto Nazionale di Astro Fisica,  via Gobetti 101, 40129 Bologna, Italy
\and Gran Sasso Science Institute (GSSI), Viale Francesco Crispi 7, 67100 L’Aquila, Italy
\and INFN-Laboratori Nazionali del Gran Sasso (LNGS), via G. Acitelli 22, 67100 Assergi (AQ), Italy}

\authorrunning{A.~Firinu et al.}
\titlerunning{The Impact of the Magnetised Cosmic Web on UHECR Propagation}
\date{Accepted ???. Received ???; in original form ???}

\abstract
{The origin of ultra-high-energy cosmic rays (UHECRs) remains an open question. Extragalactic magnetic fields can modify their propagation and, at sufficiently low energies, suppress the observed flux through the magnetic horizon (MH) effect.}
{We quantify the impact of the MH on the propagation of UHECR protons using cosmological simulations and a dedicated numerical framework that follows cosmic rays in a time-evolving background.}
{We use \texttt{UMAREL}, a parallel code developed for this study, to propagate UHECR protons through a cosmological volume simulated with ENZO. The magnetic-field configurations are chosen to be consistent with recent radio constraints on magnetic fields in cosmic-web filaments. Unlike stationary approaches, we follow particle trajectories through a sequence of time-evolving snapshots and compare the resulting arrival properties with those in an unmagnetised reference model.}
{We find that observationally motivated extragalactic magnetic fields progressively suppress the flux of arriving protons below \(E \lesssim 3 \times 10^{19}\,\mathrm{eV}\) through an effective Magnetic Horizon (MH). We estimate \(R_{\mathrm{MH}} \sim 50\,\mathrm{Mpc}\) for protons with \(E = 10^{18}\,\mathrm{eV}\) and \(R_{\mathrm{MH}} \sim 150\,\mathrm{Mpc}\) for protons with \(E = 10^{19}\,\mathrm{eV}\).}
{The MH generated by extragalactic magnetic fields must be taken into account when modelling UHECR propagation and interpreting the spectrum observed in the local Universe.}

\maketitle

\section{Introduction}
\label{sec:intro}

Ultra-high-energy cosmic rays (UHECRs) are observed up to a few $10^{20}\,\mathrm{eV}$, well beyond the energy range where Galactic magnetic fields are expected to confine charged particles efficiently~\citep{Coleman2023aph,Globus2025araa}.
Over the last decade, increasingly precise measurements by the Pierre Auger Observatory and the Telescope Array have considerably improved our knowledge of the UHECR energy spectrum, mass composition, and large-scale anisotropy~\citep{TA2018arxiv,Ivanov2019icrc,PierreAuger2019icrc,PierreAuger2020prd,PierreAuger2021epjc,TA2021apj,PierreAuger2022icrcb}. In particular, the detection by the Pierre Auger Observatory of a significant large-scale dipole above $8\times10^{18}\,\mathrm{eV}$ provides compelling evidence that the dominant UHECR population in this regime is extragalactic~\citep{PierreAuger2017science,PierreAuger2022icrc}.

Despite this progress, identifying the sources and acceleration mechanisms of UHECRs remains challenging. The difficulty is that the signal observed at Earth reflects not only the properties of the accelerators, but also the cumulative effects of propagation through radiation backgrounds and magnetic fields. At the highest energies, interactions with the cosmic microwave background (CMB) and the extragalactic background light (EBL) impose a finite attenuation scale, giving rise to the well-known Greisen-Zatsepin-Kuzmin (GZK) suppression~\citep{Greisen1966prl,Zatsepin1966jetpl}. At the same time, joint fits of the observed spectrum and composition indicate that the UHECR population above the ankle is consistent with a mixed nuclear composition and a maximum acceleration rigidity of only a few EV~\citep{PierreAuger2017jcap,PierreAuger2024jcap}. However, if one assumes a simple rigidity-dependent power-law injection spectra ($N_{inj}(E) \propto E^{-\gamma}$) at the sources and unmagnetised propagation, these fits often favour extremely hard source spectra, with injection indices $\gamma \lesssim 1$ and in some cases even $\gamma < 0$~\citep{PierreAuger2017jcap,PierreAuger2023jcap}. Such values are difficult to reconcile with standard acceleration scenarios, which more naturally predict softer spectra with $\gamma \gtrsim 2$. This tension suggests that the spectrum injected by the sources is likely modified either within the source environment or during propagation to Earth.

A particularly appealing propagation-based explanation is provided by extragalactic magnetic fields (EGMFs)~\citep{Mollerach2025uhecr}. Since magnetic deflections depend on particle rigidity, EGMFs can increase the effective path length of UHECRs, isotropise their arrival directions, and suppress the contribution of distant sources at sufficiently low rigidity. When the propagation time becomes comparable to the age of the Universe, particles from beyond a finite distance can no longer reach the observer. This defines the so-called Magnetic Horizon (MH)~\citep{Aloisio2004apj,Lemoine2005prd}. The MH can reshape the observed spectrum below the regime dominated purely by photo-hadronic losses, modify the relative contribution of different nuclear species, and affect the expected anisotropy pattern. In particular, recent analyses by the Pierre Auger Collaboration have shown that including an MH-like suppression can reconcile the data with significantly softer source spectra than those inferred in unmagnetised models~\citep{Gonzalez2021prd,PierreAuger2024jcap}.

The main difficulty, however, is that the MH depends sensitively on the strength, coherence scale, and topology of the EGMF. Most semi-analytic descriptions are based on homogeneous turbulent magnetic fields and are therefore useful mainly as idealised benchmarks. Real UHECR propagation instead takes place through the highly inhomogeneous magnetic environment of the cosmic web, where clusters, filaments, sheets, and voids contribute very differently to particle transport. In such a setting, the MH is expected to depend not only on the overall magnetic-field amplitude, but also on the geometry and intermittency of the magnetised structures crossed by the particles, as well as on the spatial distribution of the sources. Cosmological simulations have indeed shown that the predicted UHECR deflections and horizon scales can vary substantially depending on the assumed origin and evolution of cosmic magnetic fields and on the adopted source model~\citep{Hackstein2016mnras,Hackstein2018mnras}. For this reason, realistic estimates of the MH require propagation calculations performed in cosmological volumes where the magnetic field is evolved self-consistently and constrained, as much as possible, by current observations.

The observational case for widespread magnetisation of the cosmic web has strengthened significantly in recent years. Statistical detections of synchrotron emission from galaxy filaments~\citep{vern21,vern23}, Faraday-rotation constraints from LOFAR~\citep{os20}, measurements in galaxy groups and surrounding large-scale structures~\citep{Anderson2024mnras}, and recent modelling of intergalactic rotation measures~\citep{Mtchedlidze2024apj} all suggest that magnetic fields are not confined to virialised haloes alone. Joint interpretations of these data with cosmological MHD simulations indicate that volume-filling primordial seed fields at the sub-nG level remain viable, and in some cases even preferred, explanations of the observed radio properties of the cosmic web~\citep{cava25,Vazza2025aa}. In this work, we adopt one such observationally motivated scenario, based on a primordial magnetic power spectrum $P_B(k)\propto k^{-1}$ and a comoving amplitude $B_{\rm rms}=0.37\,\mathrm{nG}$.

Our goal is to quantify how these realistic cosmic-web magnetic fields affect the propagation of UHECR protons and to measure the corresponding magnetic horizon. To this end, we combine a recent cosmological MHD simulation performed with \texttt{ENZO} with the new parallel code \texttt{UMAREL}, specifically developed to follow charged particles in a time-dependent cosmological environment. Unlike stationary propagation approaches, our framework evolves UHECRs through a sequence of cosmological snapshots, thereby accounting simultaneously for the growth of large-scale structure, the redshift dependence of energy losses, and the progressive magnetisation of the cosmic web driven by both primordial fields and astrophysical feedback. In this way, we aim to assess how much observationally motivated EGMFs can reduce the distance from which UHE protons can reach observers in the local Universe, and to compare this with the corresponding unmagnetised case.

This paper is organised as follows. In Sect.~\ref{sec:methods} we describe the \texttt{ENZO} simulation and the \texttt{UMAREL} propagation framework. In Sect.~\ref{sec:results} we present the resulting particle spectra and our measurement of the magnetic horizon. In Sect.~\ref{sec:discussion} we discuss the implications and limitations of our modelling, and in Sect.~\ref{sec:conclusions} we summarise our conclusions.

\section{Methods}
\label{sec:methods}

\subsection{ENZO cosmological simulations}
\label{sec:enzo}

We use a recent cosmological magnetohydrodynamical simulation performed with the ENZO code\footnote{\url{enzo-project.org}} and described in detail by ~\citet{Vazza2025aa}. The simulated volume spans $42.5^3~\mathrm{Mpc}^3$ (comoving) and is sampled with a uniform grid of $1024^3$ cells, corresponding to a constant spatial resolution of $41.5$ kpc\,cell$^{-1}$ and a dark-matter particle mass of $1.01 \times 10^{7}\,\mathrm{M_{\odot}}$. The run includes equilibrium gas cooling and a subgrid dynamo amplification model at runtime, while the prescriptions for star formation, stellar feedback, and AGN feedback were calibrated to reproduce a number of key observable properties of the galaxy distribution.

We adopt a specific stochastic model for the primordial magnetic field, in which the magnetic-field modes are drawn from a power-law spectrum with $P_B(k)\,dk = P_{B0}k^{-1}\,dk$ and normalisation $B_{\rm rms}= 0.37 \,\mathrm{nG}$, as explained in Sect.~\ref{sec:intro}. The ENZO simulation then evolves this field self-consistently from $z=30$ to $z=0$ by solving the MHD equations. The initial and evolved magnetic power spectra of the simulation are shown in Fig.~\ref{fig:spectra}; these are discussed later in the paper in comparison with other magnetic spectra. During cosmic evolution, stellar and AGN feedback also inject magnetic fields into cells, leading to magnetised bubbles associated with matter haloes. Each feedback event is linked to the matter accretion measured at runtime in the simulation and, in the adopted model, channels $10\%$ of the total feedback energy into magnetic energy. A more detailed description of the feedback implementation, its calibration, and the other simulation parameters, which are not directly relevant here, is given in \citet{Vazza2025aa}.

\begin{figure*}[ht]
\includegraphics[width=0.995\textwidth]{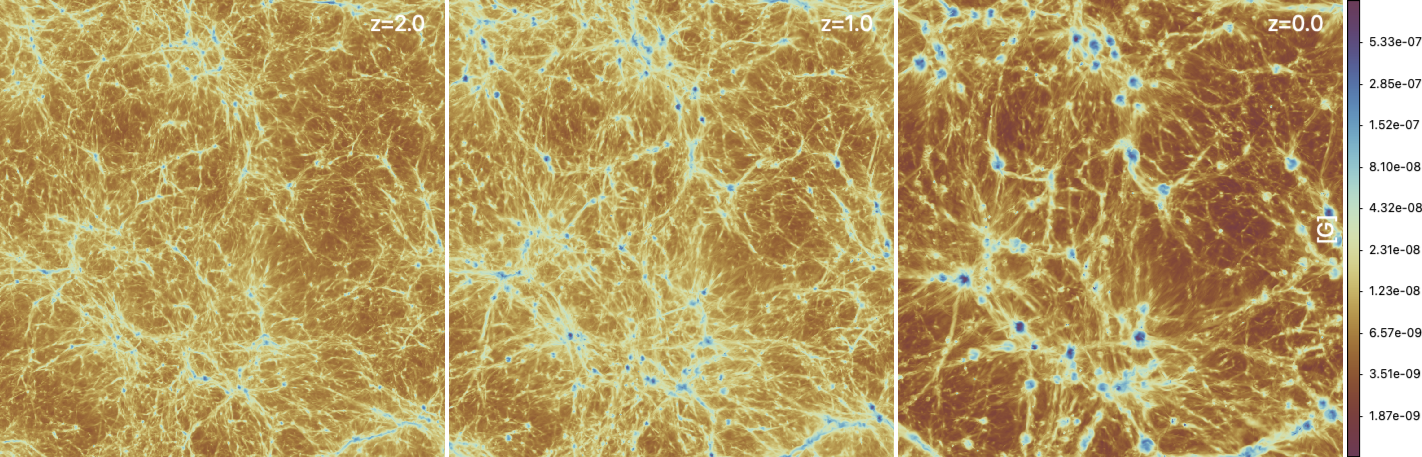}
    \caption{Projection of the mean (mass-weighted) magnetic field strength along the line of sight of the simulation at three different redshifts used in our analysis ($z=2.0, 1.0, 0.0$).}
    \label{fig:sim}
\end{figure*}


Figure \ref{fig:sim} shows the evolving projected distribution of  magnetic fields along the line of sight in the simulation at three different redshifts ($z=2.0, 1.0, 0.0$), providing a representative view of the environment traversed by our UHECRs during propagation. To first order, EGMFs trace the cosmic web, and the regions with $B\geq 10^{-9}$ G (yellow/orange) are concentrated mainly along filaments and around matter haloes. In most haloes, the magnetic field is further amplified by the subgrid dynamo model and, especially, by feedback-driven injection, reaching $B\geq 10^{-6}$ G. Nevertheless, these strongly magnetised regions are highly localised, and their volume-filling factor is therefore low. In cosmic voids, by contrast, the magnetic field remains below $10^{-9}$ G and is not significantly affected by feedback-driven magnetisation, which by $z=0$ affects only $\leq 15\%$ of the cosmic volume~\citep{Vazza2025aa}. Our main motivation for developing {\texttt{UMAREL}}, as detailed in the next section, was to follow the complex, time-dependent superposition of the volume-filling PMF and the inside-out magnetisation driven by feedback as galaxies evolve, while UHECRs propagate across the cosmic web.

\subsection{UMAREL}
\label{umarel}

The analysis of UHECR propagation starts with the injection of particles into the simulated volume described in Sect.~\ref{sec:enzo}. This is done with the parallel Julia-based numerical code \texttt{UMAREL}\footnote{\url{https://en.wikipedia.org/wiki/Umarell}} (\textbf{U}ltra-high-energy cosmic rays in \textbf{M}agnetic fields \textbf{A}ffected by \textbf{R}igidity diffusion and \textbf{E}nergy \textbf{L}osses), which injects large sets of cosmic rays into the simulated volume and self-consistently evolves their spatial trajectories and energies in time\footnote{\url{https://github.com/FrancoVazza/UMAREL_PUBLIC/}}.

In the standard setup, cosmic rays in \texttt{UMAREL} are injected only at the centres of matter haloes with 
$M_{100}\geq 10^{12}\,M_{\odot}$, in order to mimic a plausible scenario in which energetic events within galaxies produce UHECRs.
The particle species can be selected among protons, helium, nitrogen, and iron nuclei, with random initial velocity vectors (in all cases assuming $|\vec{v}| = c$) and initial energies randomly drawn in the range $10^{17}$--$10^{22}\,\mathrm{eV}$. For this work, however, we focus only on the injection and propagation of UHE protons. The initial energy is sampled uniformly in $\log(E)$ space, which is equivalent to generating an injection spectrum $N(E) \propto E^{-1}$. 
The reason for choosing this slope is to generate a sufficient number of trajectories at both the low- and high-energy ends of the spectrum, so as to obtain a similar number of particles per decade across the full energy range considered.
In order to obtain a more realistic representation, we can appropriately weight, in post-processing, the simulated UHECR statistics as a function of energy, so that the injection spectrum behaves as $N(E) \propto E^{-2}$, which represents a commonly adopted benchmark case.
 
The particle propagator implemented in \texttt{UMAREL} is based on the Boris pusher method \citep{Decyk2023cophc}, which solves the relativistic equations of motion for charged particles moving at $|\vec{v}|=c$ under the influence of the magnetic field and under periodic boundary conditions. 
This computational method was adopted because its time-centred leapfrog structure provides good numerical stability and robust long-term conservation properties in charged-particle dynamics.

\begin{figure}[ht]
\centering
\includegraphics[width=0.49\textwidth, trim = {1cm 0 0 0}, clip]{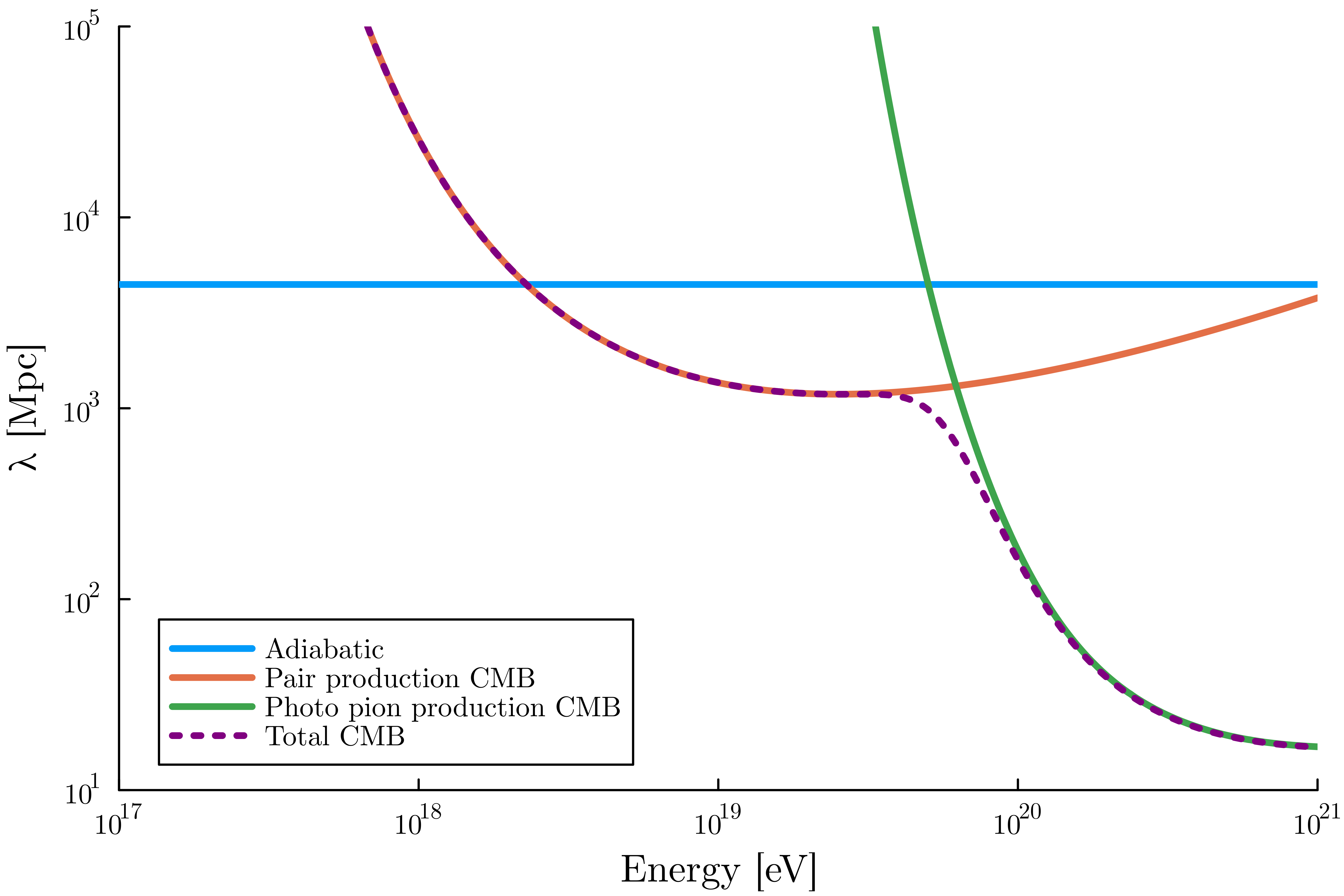}
\caption{Energy-loss lengths for high-energy protons interacting with the CMB radiation field through pair production (orange curve) and photopion production (green curve), and affected by the adiabatic expansion of the Universe (blue horizontal line). The purple dashed curve shows the sum of the two CMB interaction processes.}
\label{fig:cdedt}
\end{figure}

Between one integration timestep and the next, \texttt{UMAREL} computes the effects of energy losses on UHECRs, which in turn affect their spatial evolution by reducing their Lorentz factor $\gamma$.
The energy-loss mechanisms considered in the code are mainly due to the expansion of the Universe and to the interaction of cosmic rays with CMB photons, with a subdominant contribution from EBL photons. These processes are implemented using tabulated loss-rate curves obtained from the public Simprop-v2r4 code~\citep{Aloisio2017jcap}\footnote{\url{https://github.com/SimProp/SimProp-v2r4}} that depend on the particle mass $A$, with a characteristic energy-loss timescale defined as $\Delta t_{\rm loss}=E/|dE/dt|$. The loss rates are computed by taking into account the appropriate redshift at which particles are found in each snapshot of the simulation, according to
\[
\frac{1}{E} \frac{dE}{dt} = - \left[ H(z) + \beta_{\gamma,0}(E(1+z)) (1 + z)^3 \right] ,
\]
where $H(z)$ represents the adiabatic energy-loss rate due to the expansion of the Universe and $\beta_{\gamma, 0}$ is the sum of the loss rates related to photopion production and pair production due to the interaction with the CMB (which is the dominant background at all epochs considered here).  The loss terms are interpolated in energy at every time step, and a fourth-order Runge-Kutta integration is used to evolve the energy of each particle. As an example, Fig.~\ref{fig:cdedt} shows the energy-loss length $\lambda\,\mathrm{[Mpc]} = c\left|\frac{1}{E}\frac{dE}{dt}\right|^{-1}$ of a proton as a function of energy for the different mechanisms responsible for the losses. The blue horizontal line corresponds to adiabatic expansion, the orange and green curves describe $\lambda$ associated with pair production and photopion production, respectively, and the dashed curve gives the combined loss length of these latter two mechanisms.

A relevant novelty of our approach is that in \texttt{UMAREL} we directly take into account the cosmological evolution of sources and magnetic fields by running the code over the sequence of snapshots produced by the cosmological simulation.
This means that, as the evolution of UHECRs is computed, we update the underlying physical fields across redshift intervals using $10$ snapshots nearly equally spaced in cosmic time.
This allows us to take into account the actual evolution of cosmological fields, both by including the $(1+z)^2$ increase of the physical magnetic field and by considering a realistic model for the evolution of astrophysical sources of magnetisation (e.g.~galaxies and active galactic nuclei), whose evolution roughly follows the cosmic star-formation history, as shown in \citet{Vazza2025aa}. 
New UHECR protons are introduced in the simulation at different epochs and at regular time intervals, as different ``generations'' ($N_{\rm gen}=25$ for the runs considered in this work), based on the halo distribution at the corresponding redshift; hence the elapsed time between two injections is $\Delta t_{\rm inj} \approx 0.4 \,\mathrm{Gyr}$ for the assumed cosmology.
In the baseline setup, we limited the injection of UHECRs to haloes with $M_{100} \geq 10^{12}\,M_{\odot}$, and considered an equal number of UHECRs for each generation event. In Appendix~\ref{A1} we provide an example of the projected distribution of selected sources in one of our snapshots, together with the volumetric distribution of magnetic fields at the same epoch. Our production runs propagated up to $\sim 10^5$ protons through $10$ different snapshots of our ENZO simulation, sampling the evolution of the Universe from $z=2.0$ to $z=0.0$. 
To the best of our knowledge, the possibility of including redshift-dependent effects and using several evolving snapshots during the simulated evolution of UHECRs is a distinctive feature of \texttt{UMAREL}, and it motivated us to develop this tool for our study of the MH.
Typical \texttt{UMAREL} production runs were performed either on a laptop ($16$ tasks) or using up to $128$ tasks on four nodes of the Leonardo supercomputer at CINECA (Italy).

\begin{figure*}[ht]
\includegraphics[width=0.48\textwidth]{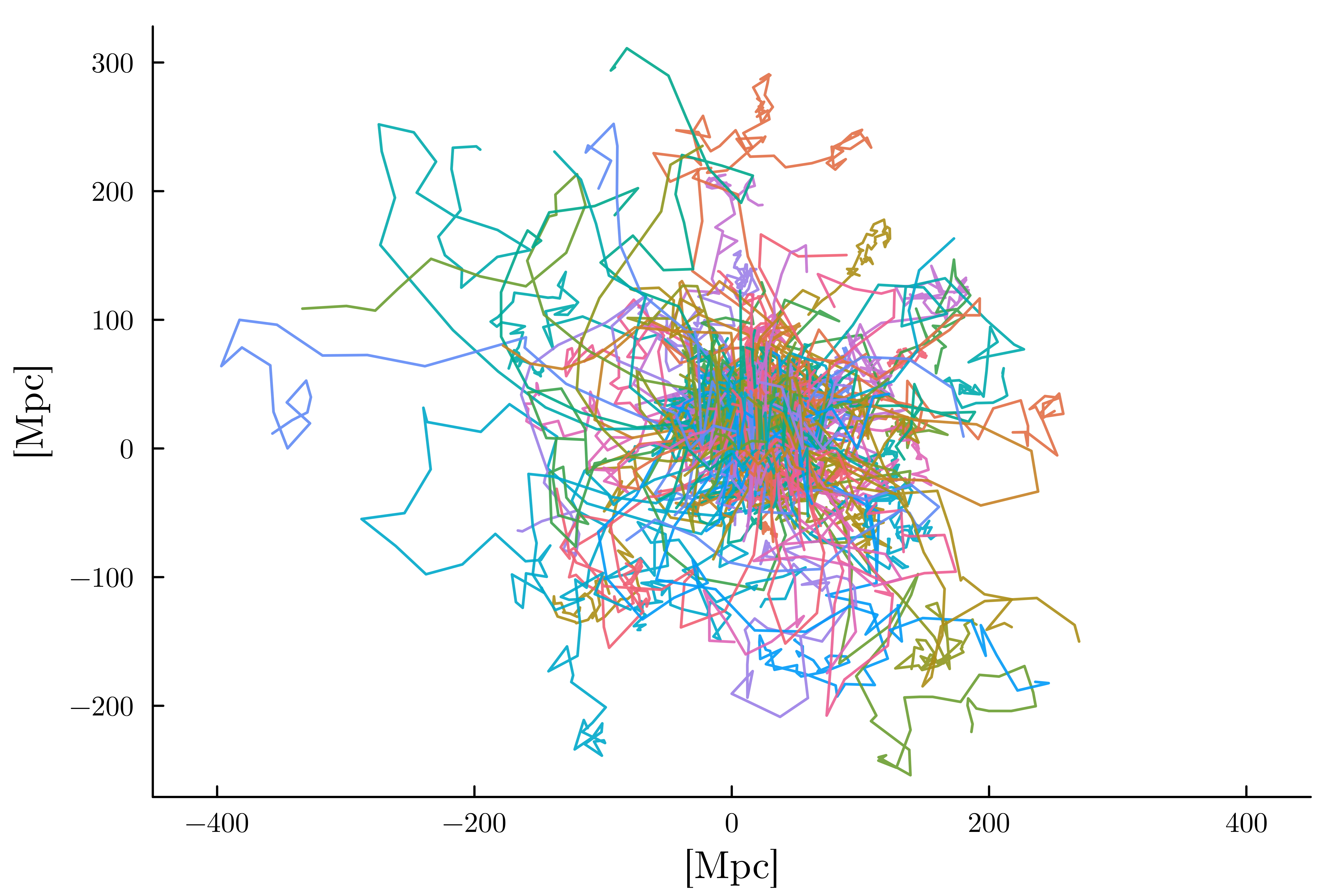}
\includegraphics[width=0.5\textwidth]{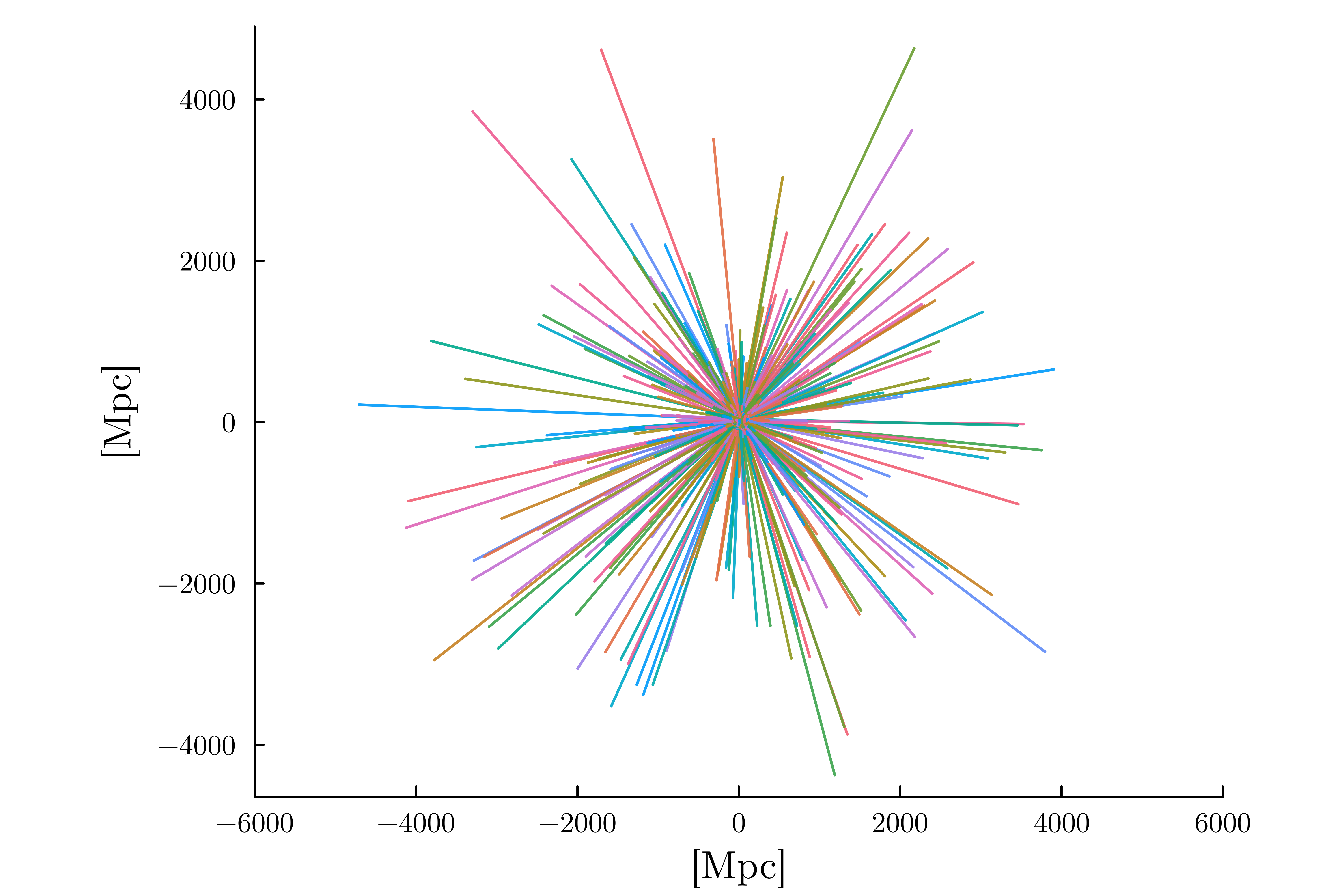}
\caption{On the left, a representation of the paths travelled by protons affected by energy losses and propagating through our baseline model for extragalactic magnetic fields. On the right, the same protons are propagated through an unmagnetised volume and are affected only by energy losses. Note the different length scales corresponding to the distances covered by particles; in both cases we assumed periodic boundary conditions, and hence the full trajectories displayed here cover a much larger distance than the simulated box size.}
\label{mapfield}
\end{figure*}

Figure \ref{mapfield} shows an example of the propagation of UHECRs evolved from $z=2$ to $z=0$ across all snapshots of the simulation and injected $N_{\rm gen}=25$ times in the energy range $10^{18}$--$10^{21}\,\mathrm{eV}$. As expected, in the case without magnetic fields (right panel), UHECRs propagate straightly at the speed of light and their propagation is limited only by energy losses; in the case of our realistic EGMF model (left panel), instead, their maximum propagation distance is considerably limited by the MH. 

In the Appendix we report several tests conducted to check the physical validity of \texttt{UMAREL} and the behaviour of its various modules (e.g.~deflection by magnetic fields, energy losses, and different strategies to select UHECR sources).

\section{Results}
\label{sec:results}

\begin{figure}[ht]
    \centering
    \includegraphics[width=1\linewidth]{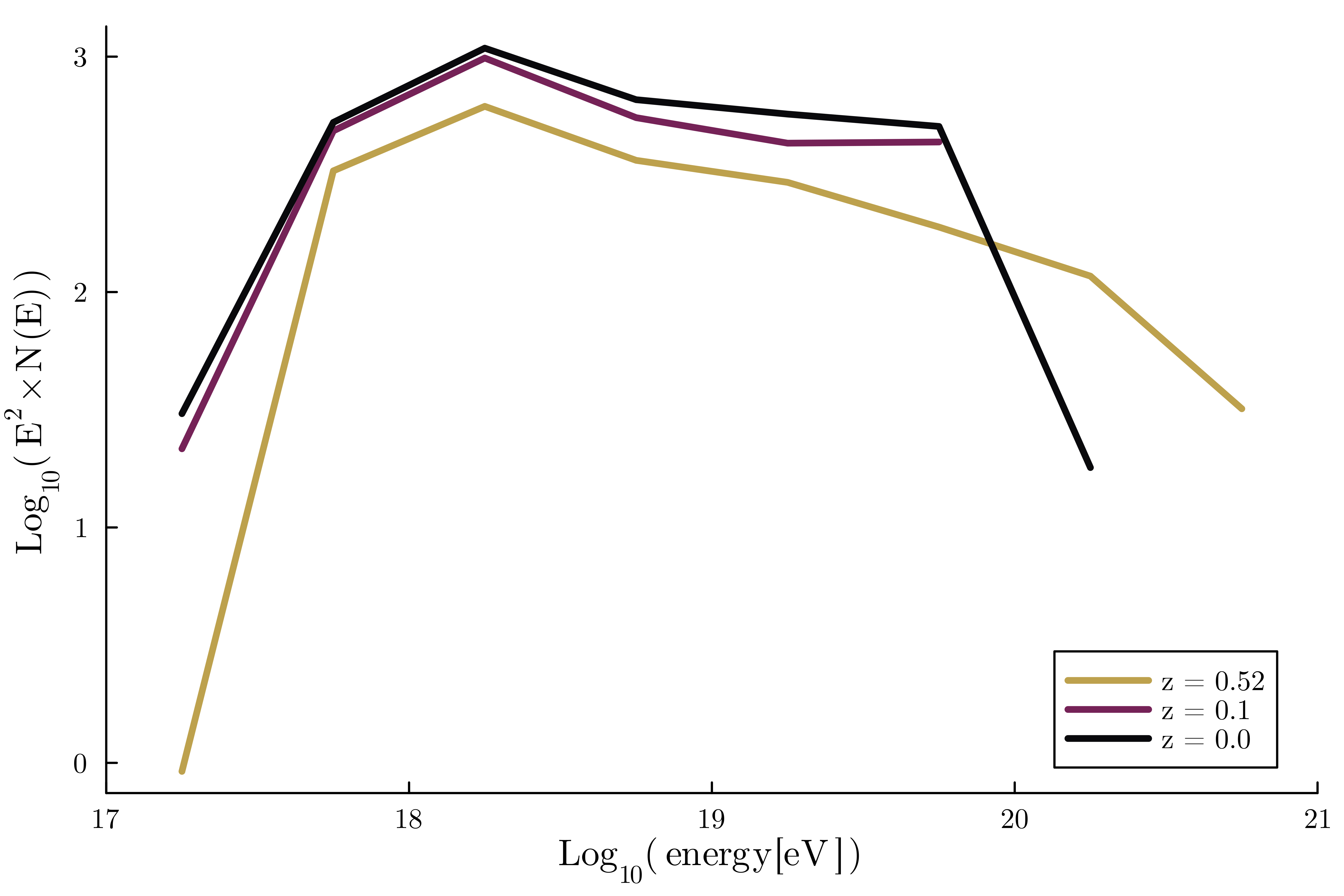}
    \caption{Energy spectra of all simulated UHECR protons collected over the full volume of the simulated Universe at $z = 0.5$, $0.1$, and $0$. Differences between the baseline magnetised model and the unmagnetised case are not visible because the spectra are virtually identical at all epochs.}
    \label{spectrum}
\end{figure}

As a first result of our analysis, in Fig.~\ref{spectrum} we show the energy spectrum of all propagated UHECR protons within our simulated volume at three different late epochs, namely $z=0.5$, $z=0.1$, and $z=0.0$. At all injection epochs, the injected particles are weighted so as to mimic a source spectrum proportional to $E^{-2}$, which would correspond to horizontal lines in the figure.

Although this cannot be appreciated visually in the plot, the spectra obtained in the baseline magnetised model and in the unmagnetised case are virtually identical. This is because magnetic fields affect the trajectories of the particles, but not their energy losses as a function of time.

At $z=0$, the total spectrum is reasonably consistent with expectations, showing a gradual steepening between $10^{18}$ and $10^{20}\,\rm eV$ due to pair production and photopion production, followed by a marked suppression above $E \geq 10^{20}\,\rm eV$. This behaviour is consistent with the classical GZK cutoff, and reflects the fact that the last of the 25 UHECR injection episodes in the simulation occurs at $z=0.02$, corresponding to approximately $300\,\rm Myr$ before the present epoch. At this low redshift of injection, the observed energy drop below $\sim 10^{18} \rm eV$ is due to the fact that we inject protons only starting from $10^{18} \rm eV$, hence energy losses did not have time to lower the particle energy below this threshold before the end of the simulation.

Of course, the equivalence between the UHECR spectra in the magnetised and unmagnetised cases holds only when all particles within the simulated volume are included. In contrast, magnetic fields do affect the spectrum measured by individual observers located at finite distances from the sources, through the magnetic-horizon effect, which we quantify next.

In order to quantify the effect of the MH in our simulation, we measure the distribution of distances travelled by protons from their injection sites to their final positions at $z=0$, where the observer is placed. Periodic boundary conditions are assumed, so that particles can in principle travel distances much larger than the size of the original simulation box.

\begin{figure}[ht]
    \centering
    \includegraphics[width=1\linewidth]{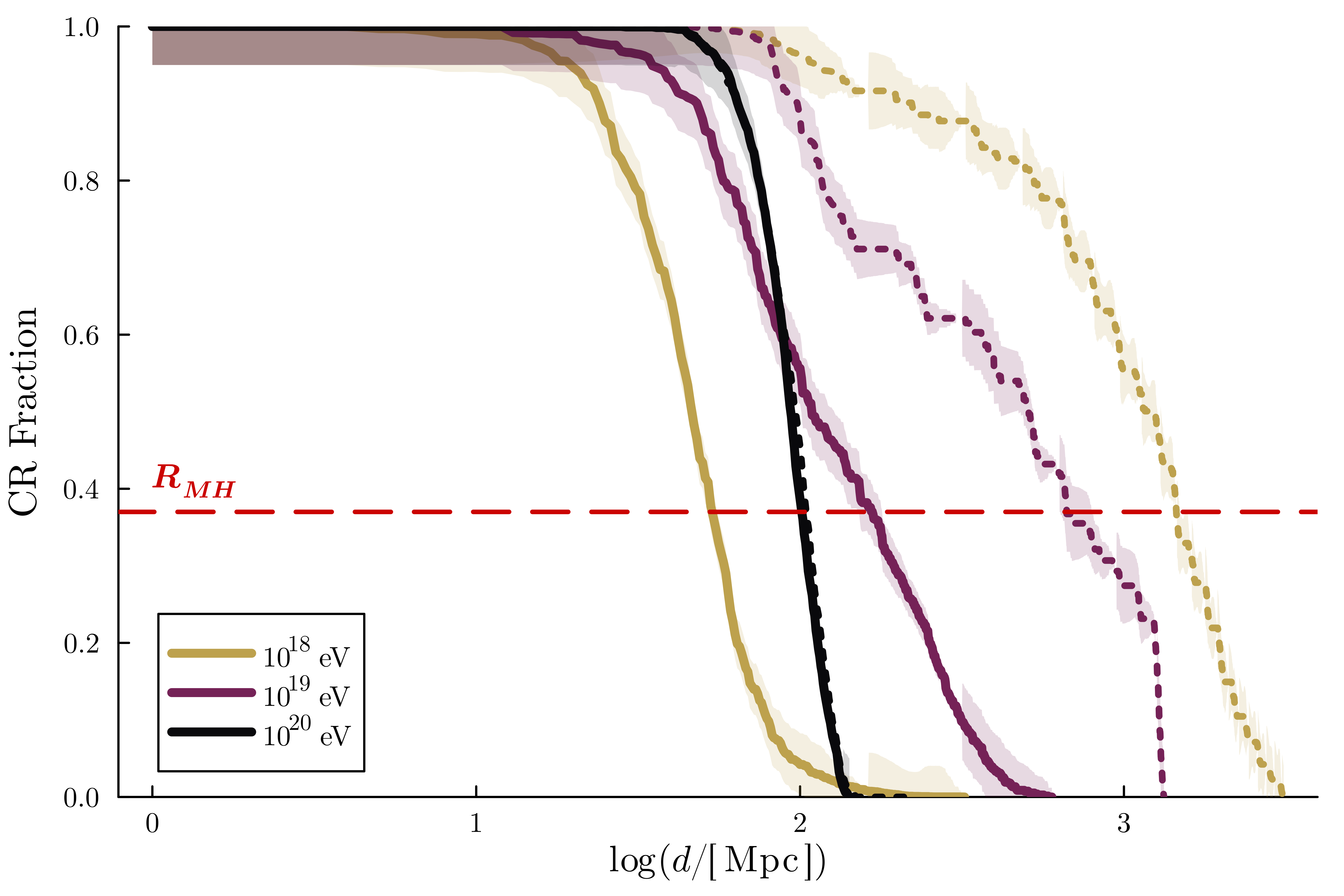}
    \caption{Distribution of the travelled distance covered by $z=0$ by protons with different ranges of arrival energies (normalised to the total number of protons within the same energy bin) for the fiducial propagation case with realistic magnetic fields  (solid curves) and in the scenario without magnetic fields (dotted curves). The shaded areas represent the standard deviation around the mean distributions in all cases. The red horizontal line marks the $37\%$ fraction we used to quantify the extent of the MH in all cases.}
    \label{fig:fidvsnofield}
\end{figure}

Figure \ref{fig:fidvsnofield} shows the fractional distribution of travelled distances for all UHECR protons as a function of their arrival energy at $z=0$, comparing propagation in realistic magnetic fields (solid curves) with the unmagnetised case (dotted curves). Each distribution is normalised to the total number of UHECRs within the corresponding arrival-energy bin. This figure clearly illustrates the impact of realistic EGMFs on proton propagation.

Protons arriving with energies of $10^{20}\,\rm eV$ (black curves) can reach the observer from distances of order $\sim 100\,\rm Mpc$, almost independently of the presence of magnetic fields, since their propagation is primarily limited by energy losses through the GZK effect. On the other hand, the differences between the magnetised and unmagnetised scenarios become increasingly significant at lower energies. Because of their lower rigidity, protons arriving at $10^{18}\,\rm eV$ (yellow curves) and $10^{19}\,\rm eV$ (purple curves) follow progressively more tangled trajectories and, on average, can travel shorter effective distances from their sources than in the unmagnetised case.

In particular, for a realistic model of EGMFs, protons in the $\sim 10^{18}\,\rm eV$ range cannot reach observers from distances much larger than $\sim 100\,\rm Mpc$. This defines an effective MH, with $R_{\rm MH} \sim 60\,\rm Mpc$ at $10^{18}\,\rm eV$ and $R_{\rm MH} \sim 120\,\rm Mpc$ at $10^{19}\,\rm eV$.

More quantitatively, we parametrise the survival fraction of particles arriving within a given energy bin by modelling it with an exponential law: 
\begin{equation}
    N(x)=N_0\,e^{-\lambda/x},
\end{equation}
where $N_0$ is the total number of protons, $x$ is the travelled distance, and $\lambda$ is the attenuation scale. In this parametrisation, $x=\lambda$ defines the characteristic distance at which the surviving fraction of protons falls to $1/e \approx 37\%$. This exponential suppression captures the combined effects of energy losses and magnetic deflections during propagation, and provides the key quantity that we use to define the MH, $R_{\rm MH}$, in our model.

\begin{figure}[ht]
    \centering
    \includegraphics[width=1\linewidth]{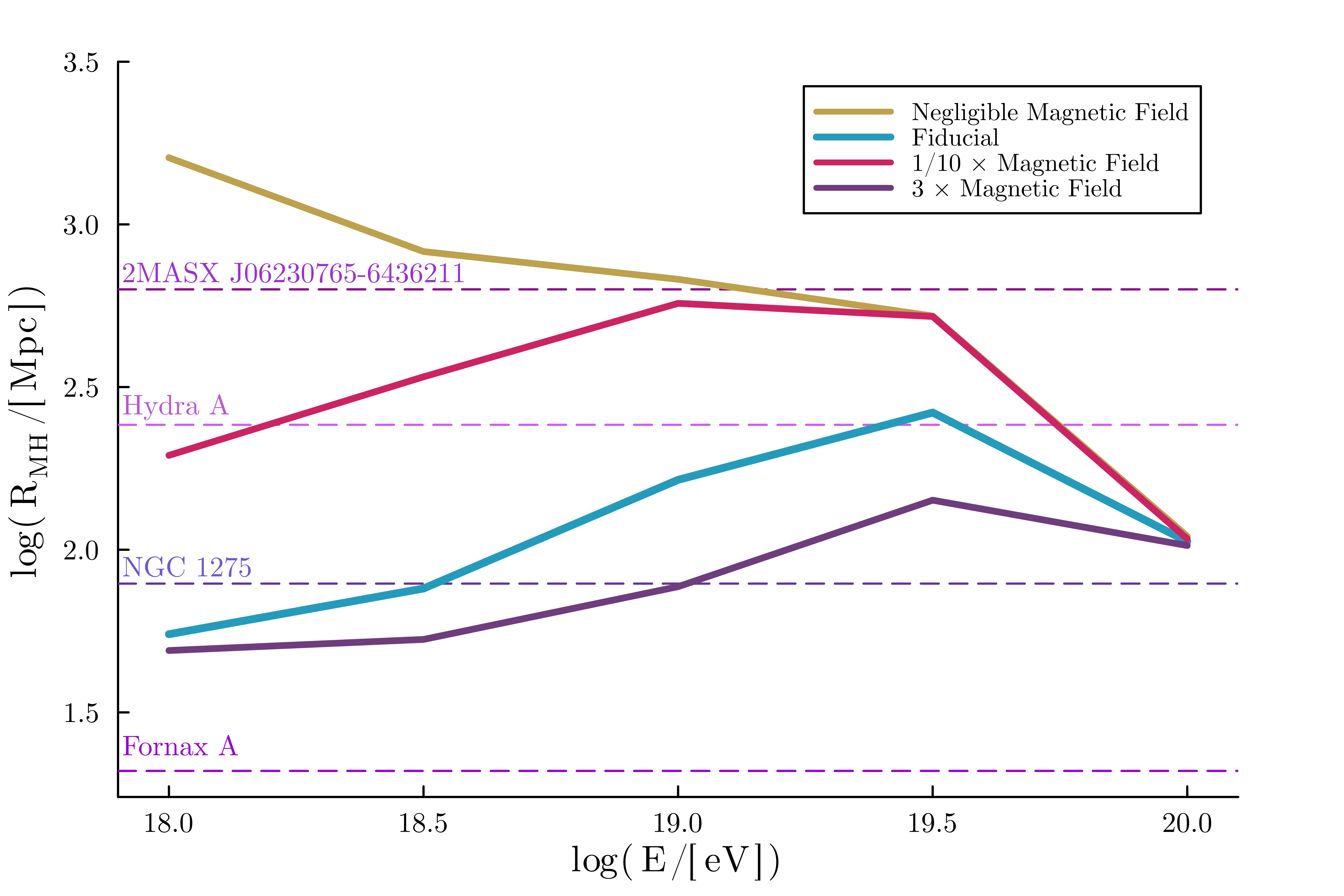}
    \caption{The MH as a function of energy and for different models of magnetic fields. The different lines give the propagation distance of $\sim63\%$ from their injection points for cosmic ray protons, as function of their  their arrival energy  at $z=0$. The curves represent the fiducial model (blue) and the negligible magnetic field scenario (yellow). The pink and purple curves refers to a propagation in an environment with a magnetic field 10 times weaker and 3 times stronger, respectively, than in the fiducial case. The horizontal lines mark the distance of possible real sources for UHECRs.}
    \label{fig:moneyplot}
\end{figure}

Figure \ref{fig:moneyplot} shows the measured trend of $R_{\rm MH}$ as a function of particle arrival energy for the fiducial magnetised model (blue curve) and for the unmagnetised case (yellow curve).

The horizontal dashed lines mark the distances of a few powerful radio galaxies in the local Universe, selected from the catalogue described by~\citet{vanVelzen2012aa}. These sources were chosen to have synchrotron power $P_{\rm syn} \geq 10^{34}\,\rm W$ at $1400\,\rm MHz$. This comparison makes it possible to estimate the low-energy cutoff expected in the observed spectrum, depending on the distance to the nearest source.

In our baseline model, the MH effectively excludes several nearby radio galaxies as plausible sources of the observed UHECR flux, at least under the assumption of a proton-dominated composition. This assumption, however, becomes progressively less justified at the highest energies, where observations indicate an increasingly heavier composition~\citep{PierreAuger2024jcap}.

We note that the MH derived in this way is defined with respect to all possible observers within the simulated volume, rather than for a specific observer, such as one located at the centre of a Milky Way-like galaxy or, more generally, within galaxies in the simulation. We adopt this definition of $R_{\rm MH}$ in order to maximise the available statistics and minimise numerical noise.
We also extensively tested an alternative approach in which only UHECRs reaching specific observers in the simulation are collected, and the MH is measured using only those events. This is a standard procedure, which requires defining a finite search radius around each observer, and is commonly adopted in constrained simulations of the local Universe \citep[e.g.][]{Hackstein2016mnras}. However, given the number of simulated UHECRs and the limited volume of our simulation, this method was found to introduce excessive numerical noise. This is mainly due to the very small volume ratio between a plausible observer sphere ($\sim \rm Mpc^3$) and the total volume over which most UHECRs are distributed ($\sim 42^3\,\rm Mpc^3$).
We therefore defer to future work with \texttt{UMAREL} the analysis of updated Local-Universe simulations including realistic UHECR observers.

\section{Discussion}
\label{sec:discussion}

\subsection{Dependence of the MH on magnetic-field strength}

Having established that neglecting magnetic fields strongly affects the maximum distance that protons can travel, we next investigate how $R_{\rm MH}$ depends on magnetic-field strength. To this end, we constructed two additional magnetic configurations by simply renormalising the intensity of the fiducial magnetic field obtained in the simulation.

The results of this analysis are also shown in Fig.~\ref{fig:moneyplot}. In addition to the comparison between the fiducial model (blue curve) and the unmagnetised case (yellow curve) discussed in Sect.~\ref{sec:results}, we also show $R_{\rm MH}$ for UHECR protons propagated through a magnetic field that is three times stronger ($\approx 1\,\mathrm{nG}$) and ten times weaker ($\approx 0.037\,\mathrm{nG}$) than in the fiducial model, shown by the purple and pink curves, respectively. In all cases, the magnetic field was rescaled in post-processing in every cell of the baseline simulation by the appropriate factor. This simplified approach is motivated by the fact that a full MHD re-simulation is expected to yield results close to those obtained through a post-processing rescaling \citep[see e.g.][]{Mtchedlidze2024apj}. Moreover, the regions affected by feedback-driven magnetisation bubbles, for which a rigid rescaling is less accurate because their magnetic field depends on the properties of the host galaxies, occupy only a small fraction of the total cosmic volume.

In Table~\ref{table}, we report the values of $R_{\rm MH}$ for representative arrival energies ($10^{18}$, $10^{19}$, and $10^{20}\,\rm eV$) and for the different magnetic-field strengths considered here. This provides a more quantitative view of how the MH depends on both particle energy and magnetic-field intensity.

Figure~\ref{fig:moneyplot} is particularly instructive because it highlights the impact of magnetic-field strength on $R_{\rm MH}$. Below $\sim 10^{19}\,\rm eV$, the four curves are clearly separated, indicating that the magnetic field strongly affects the propagation of UHECR protons in this regime. As the energy increases, however, the curves become progressively more similar, and for $E \geq 10^{20}\,\rm eV$ they converge towards $R_{\rm MH} \sim 100\,\rm Mpc$.

\renewcommand{\arraystretch}{1.55}
\begin{table*}[ht]
\small
    \centering
    \begin{tabular}{|c|c|c|c|c|}
    \hline
    \textsc{Energy} & $B=0$ & $B_{\rm rms}=0.04\,\rm nG$ & {\bf $B_{\rm rms}=0.37\,\rm nG$} & $B_{\rm rms}=1.0\,\rm nG$ \\
    $\rm [\,eV]$ & \textsc{$R_{\rm MH}$ [Mpc]} & \textsc{$R_{\rm MH}$ [Mpc]} & \textsc{ $R_{\rm MH}$ [Mpc]} & \textsc{$R_{\rm MH}$ [Mpc]} \\
    \hline
    $10^{18}$ & $1458.94\pm42.95$ & $184.67\pm3.74$ & \textbf{$53.54\pm0.93$} & $48.80\pm1.31$ \\
    \hline
    $10^{19}$ & $563.47\pm24.41$ & $490.08\pm14.72$ &  \textbf{$153.56\pm4.83$}& $72.81\pm1.87$\\
    \hline
    $10^{20}$ & $103.71\pm0.58$ & $102.22\pm1.01$ & \textbf{$100.89\pm1.42$} & $99.28\pm9.04$ \\
    \hline
    \end{tabular}
    \caption{MH, $R_{\rm MH}$, estimated as the characteristic distance within which $\sim 63\%$ of cosmic-ray protons observed at a given energy have travelled by $z=0$, along with the corresponding uncertainty computed within $1\sigma$ around the mean value of $R_{\rm MH}$. These results refer to the fiducial model and the three alternative magnetic-field configurations considered here: the unmagnetised case, a field ten times weaker than fiducial, and a field three times stronger than fiducial.} 
    \label{table}
\end{table*}

It is also important to note that the values of $R_{\rm MH}$ shown in Fig.~\ref{fig:moneyplot} are not determined by magnetic deflections alone, but also reflect the role of the energy-loss processes discussed in Sect.~\ref{umarel}. At low energies, protons have lower rigidity and are therefore more strongly affected by magnetic fields, but they experience relatively mild energy losses during propagation. At high energies, by contrast, protons would in principle be able to travel larger distances because of their larger rigidity, but they are also subject to much stronger energy losses, which reduce their energy during propagation and therefore limit their effective horizon.

This explains the overall behaviour of $R_{\rm MH}$. At low energies, the four magnetic-field configurations produce clearly distinct horizons, because magnetic confinement is the dominant effect. At higher energies, instead, the different cases progressively converge, as energy losses become increasingly important and reduce the maximum propagation distance in all models.

\subsection{Tests on the magnetic field topology}
\label{subsec:spectra}

Next, we examine to what extent the complex topology of cosmic magnetic fields affects the propagation of UHECR protons by comparing the realistic distribution produced by our cosmological MHD simulation with more idealised magnetic-field configurations.

To enable a direct comparison with previous work, we constructed two additional models for the three-dimensional distribution of magnetic-field vectors, generated without evolving the field through MHD. In detail, on a $1024^3$ grid covering the same $42.5^3\,\rm Mpc^3$ volume, we generated:
(a) a three-dimensional magnetic field randomly drawn from a power-law magnetic power spectrum with slope $P_B(k)\propto k^{-1}$, where a $k^{-3}$ scaling would correspond to a scale-invariant field, and with a maximum coherence scale equal to the box size (hereafter, model K-1);  
and (b) a three-dimensional magnetic field randomly drawn from a power-law spectrum with slope $P_B(k)\propto k^{-11/3}$, corresponding to Kolmogorov scaling, and with a maximum coherence scale of $1\,\rm Mpc$ comoving (hereafter, model K-11/3).

In both cases, the rms magnetic-field strength was set to $B_{\rm rms}=\langle B^2\rangle^{1/2}=1\,\rm nG$.

\begin{figure}[ht]
\includegraphics[width=0.495\textwidth]{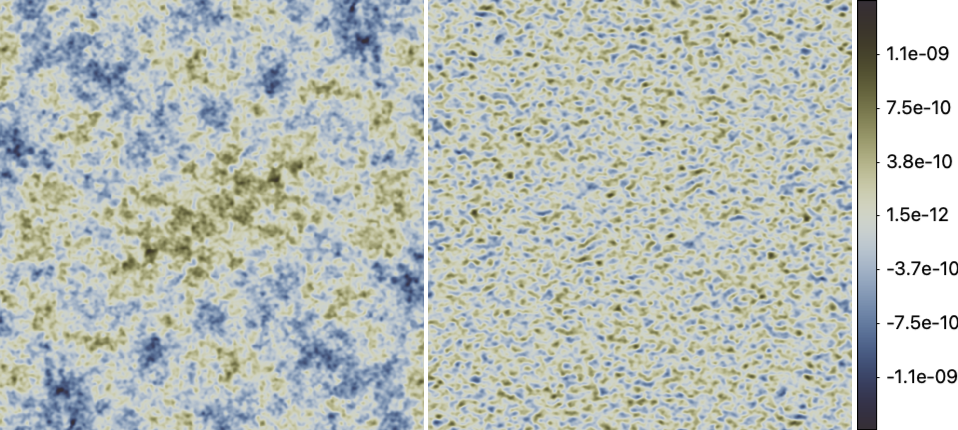}
    \caption{Slices showing the magnetic-field component along the line of sight (in units of Gauss) through a simulated $42.5^3\,\rm Mpc^3$ volume, for two idealised magnetic-field models: a stochastic field drawn from a $P_B(k)\propto k^{-1}$ spectrum of fluctuations with maximum scale equal to the computational domain (model K-1, left), and a stochastic field drawn from a Kolmogorov spectrum, $P_B(k)\propto k^{-11/3}$, with a maximum scale of $1\,\rm Mpc$ (model K-11/3, right).}
    \label{fig:maps_synthetic}
\end{figure}

The first case corresponds to the initial condition of our baseline simulation, although here placed at $z=0$, while the second is representative of the magnetic-field models commonly adopted in the UHECR propagation literature \citep[e.g.][]{Aloisio2004apj,Mollerach2025uhecr}.
Figure \ref{fig:maps_synthetic} shows the magnetic-field distribution in a thin slice through both models.

A third model used for comparison is the magnetic-field configuration adopted in our main analysis, which, for consistency with the other two cases, we renormalised in post-processing so that its rms field strength is also $1\,\rm nG$. Since in this model the global rms value is biased high by the localised contribution of feedback-driven magnetisation bubbles, the renormalisation was computed using only the rms field in the density range $\rho/\langle\rho\rangle \leq 10$, which is only weakly affected by astrophysical magnetisation.

In all three cases, using exactly the same set of sources, we studied the propagation of UHECR protons in the energy range $10^{18}$--$10^{21}\,\rm eV$ over $\approx 8\,\rm Gyr$ across the same periodic volume of $42.5^3\,\rm Mpc^3$. In order to isolate the role of magnetic-field topology, we did not include energy losses or cosmological expansion in this set of runs, and we followed a single generation of UHECR protons injected at the beginning of the run ($z=1$).

\begin{figure}[ht]
	\centering
	\includegraphics[width=1\linewidth]{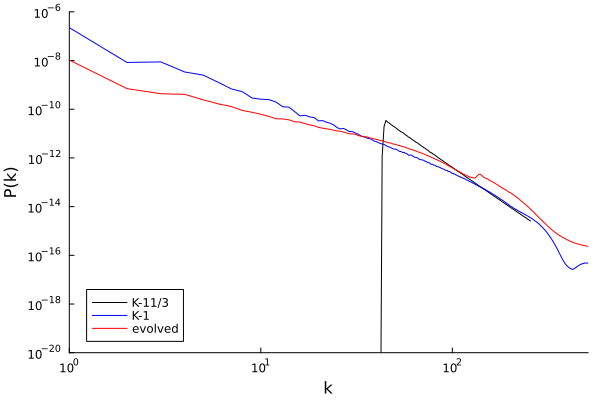}
	\caption{Magnetic-field power spectra (in units of [$\rm G^2/d^3k$], where $k=2\pi/L_{\rm box}$ is the wavenumber and $L_{\rm box}$ is the normalised box size, corresponding to $42.5\,\rm Mpc$ comoving) for the three magnetic-field configurations used in Sect.~\ref{subsec:spectra}.}
	\label{fig:spectra}
\end{figure}

Figure \ref{fig:spectra} shows the three-dimensional power spectra of the magnetic field for these three configurations. Models K-1 and its MHD-evolved counterpart are, as expected, broadly similar. However, by $z=0$ the dynamical evolution of the gas has modified the original spectrum, producing a flatter distribution and a significant excess of magnetic power on scales $\lesssim 1$--$2\,\rm Mpc$, which are the scales most strongly affected by the injection of magnetic energy through feedback events in galaxies.

\begin{figure*}[ht]
    \centering
    \includegraphics[width=1\linewidth, trim={0 6cm 0 0}, clip]{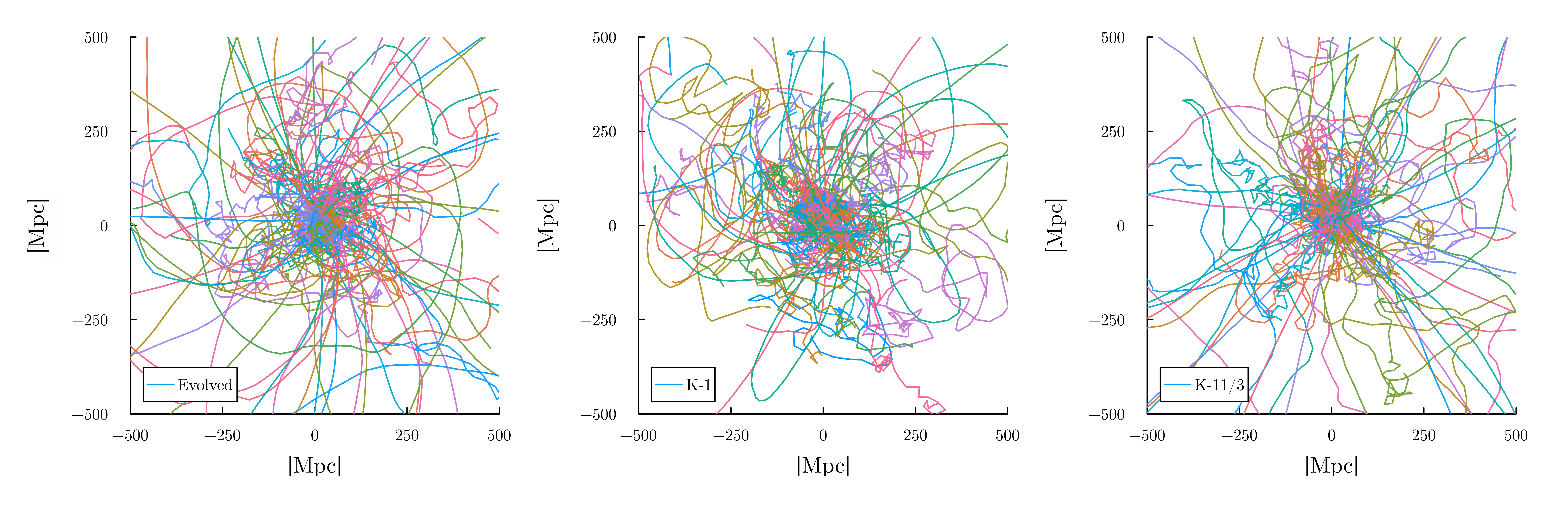}
    \caption{Representative proton trajectories in the three magnetic-field configurations, all normalised to $B_{\rm rms}=1\,\rm nG$: the evolved baseline model (left), K-1 (centre), and K-11/3 (right). In these runs, proton propagation is not affected by energy-loss processes.}
    \label{fig:mappe}
\end{figure*}

In Fig.~\ref{fig:mappe} we show the projected propagation paths of protons in the three cases, from their injection at $z=1$ to the end of the run.

\begin{figure}[ht]
    \centering
    \includegraphics[width=1\linewidth]{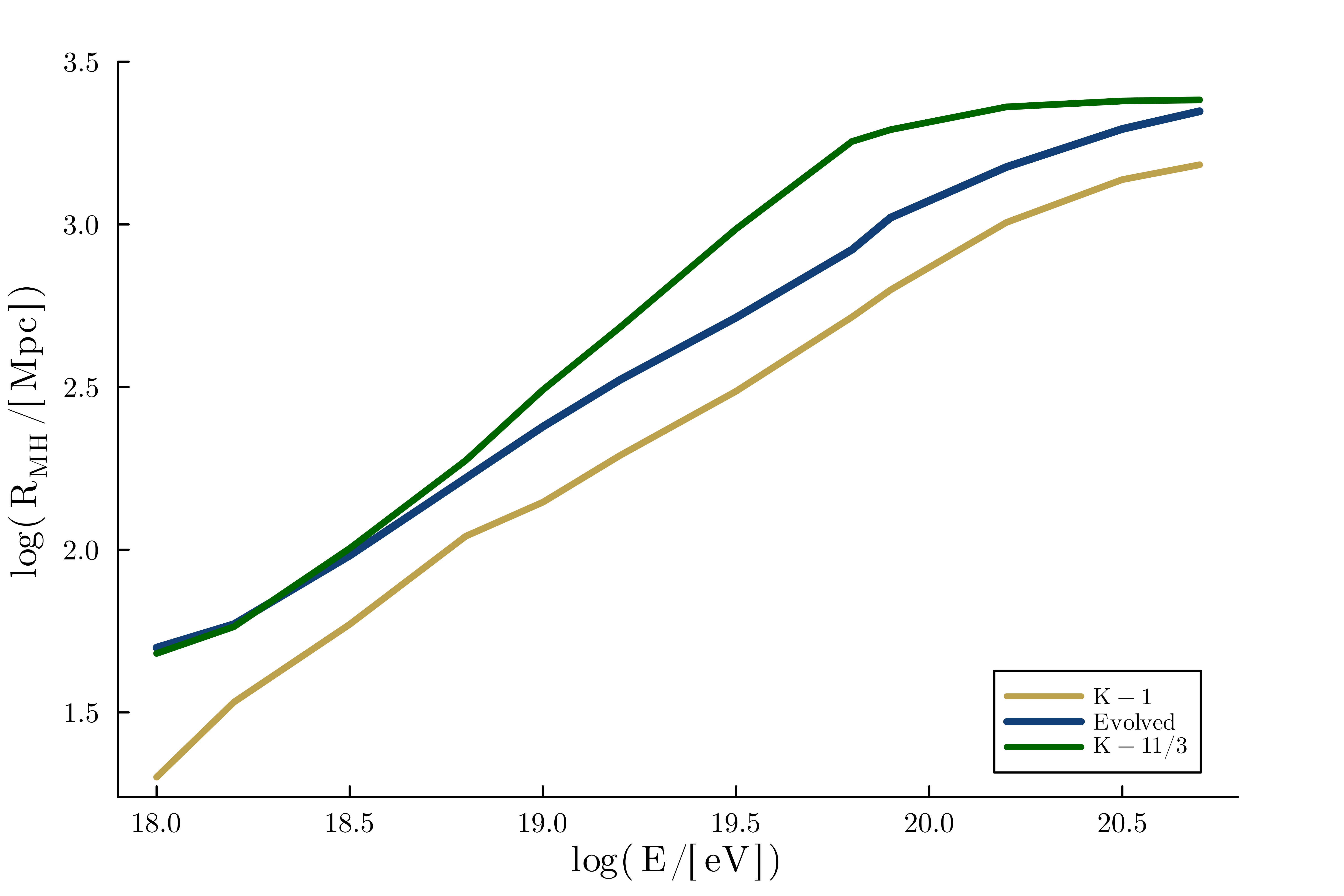}
    \caption{Maximum propagation distances within which $\sim63\%$ of cosmic ray protons are expected to be injected, as a function of their arrival energy measured at $z\sim0$. The curves represent the evolved model (blue) and the two synthetic schemes, $K-11/3$ (green) and $K-1$ (yellow), all with $B_{rms}=1$ nG.}
    \label{fig:comparisonmodel}
\end{figure}

Using the same approach adopted above, we again computed the fractional distribution of the distances travelled by protons in different energy bins, from injection to the end of the run at $z=0$. The results are shown in Fig.~\ref{fig:comparisonmodel}.

Each curve again provides the characteristic propagation distance enclosing $63\%$ of protons in the three models considered above: K-1 (yellow), K-11/3 (green), and the evolved model (blue). The impact of magnetic-field topology on proton propagation is significant, even in the absence of energy losses and even when all three models are normalised to the same rms field strength, $B_{\rm rms}=1\,\rm nG$. At all energies, the largest distances are reached by protons propagating in the K-11/3 model. This can be readily understood because the Larmor radius of a $10^{18}\,\rm eV$ proton in a $1\,\rm nG$ field is $r_{\rm L}\approx 1\,\rm Mpc$. Therefore, for the energies considered here, protons gyrate on scales comparable to or larger than the maximum coherence scale of this magnetic-field model.

By contrast, protons travel the shortest distances in the K-1 model, because in this case the magnetic field is correlated up to scales comparable to the box size and exhibits stronger fluctuations over a broad range of scales, as illustrated in Fig.~\ref{fig:maps_synthetic}. In both synthetic cases, the magnetic field is not coupled to the dynamical evolution of the large-scale structure, so the propagation of particles depends primarily on the shape of the magnetic power spectrum. Protons are therefore mainly scattered by magnetic fluctuations on scales comparable to their Larmor radii. In the K-1 model, a larger fraction of the magnetic power is distributed on the scales relevant for scattering, which leads to stronger deflections. In the K-11/3 case, the power at those same scales is lower, allowing particles to propagate more efficiently through the volume.

The propagation properties of the evolved model lie between these two limiting cases. Once the dynamical evolution of the magnetic field is taken into account, the simulated volume naturally develops both weakly magnetised regions, such as cosmic voids, where protons are less efficiently confined, and strongly magnetised environments surrounding dense structures, where propagation can instead be significantly hindered.

Overall, we find that at all energies the MH can be overestimated or underestimated by a factor of $\sim 1.5$--$2$ when idealised, volume-filling power-law magnetic-field models are used in place of the more complex magnetised cosmic-web distribution produced by cosmological MHD simulations.


\subsection{Limitations and future developments}

Assessing the limitations of our work is important in order to identify the most relevant directions for future improvements.

A first obvious limitation is the relatively small simulated volume, $(42.5\,\mathrm{Mpc})^3$, which does not capture the full variety of magnetic environments that UHECRs may encounter during propagation and may also be affected by cosmic variance. In addition, such a limited volume cannot host the formation of superclusters or of the longest filaments of the cosmic web, both of which may be relevant for UHECR transport on large scales.

A second limitation concerns the spatial resolution of the simulation, $41.5\,\mathrm{kpc}$, which may not be sufficient to fully resolve magnetic structures on the smallest scales, nor the detailed density stratification within the candidate UHECR sources.

Moreover, our present simulations do not yet provide a realistic representation of the Local Universe. This limitation can be addressed in future work by adopting constrained simulations, following the approach explored in a number of recent studies \citep[e.g.][]{Dolag2003icrc,Hackstein2018mnras,Boss2024aa}.

Finally, extending the analysis to heavier nuclei will provide a more complete view of the role of magnetic fields in shaping the full UHECR spectrum, and will also help to better constrain their origin and composition. This can be achieved either through future updates of our code \texttt{UMAREL}, or by comparison with other publicly available propagation tools \citep[e.g.][]{CRPROPA3}.

\subsection{Comparison with previous studies of the MH}

It is useful to compare our results with previous studies that investigated whether the MH effect can account for the low-energy suppression of the UHECR flux, as inferred from Pierre Auger Observatory data~\citep{PierreAuger2017jcap}. In particular, \citet{PierreAuger2024jcap} showed that, for the MH to play a major role in shaping the observed spectrum and composition, the source spacing and magnetic-field properties must satisfy rather stringent conditions. In their framework, the effect becomes relevant only for sufficiently sparse sources, corresponding to intersource distances $d_s \gtrsim 20\,\mathrm{Mpc}$ for $L_{\rm coh}\sim 100\,\mathrm{kpc}$, and for parameter combinations such that $X_s R_{\rm crit}\simeq 5$--$10\,\mathrm{EeV}$. For typical source separations of order $20\,\mathrm{Mpc}$, this translates into local extragalactic magnetic fields of order several tens of nG, typically $\sim 50$--$100\,\mathrm{nG}$ for coherence lengths of order $100\,\mathrm{kpc}$.

Our results point to a milder effect. In Fig.~\ref{fig:moneyplot}, or equivalently in the energy range shown in Fig.~7, we find that the MH in our fiducial model is of order $\sim 50$--$100\,\mathrm{Mpc}$ at $\log_{10}(E/\mathrm{eV})\simeq 18.0$--18.5. This demonstrates that a more realistic distribution of extragalactic magnetic field, dynamically coupled with the cosmic web, can already produce a non-negligible suppression of the accessible source volume at EeV energies, using a significantly lower magnetic field. However, the level of suppression that we obtain appears weaker than would be required for the MH to provide, by itself, the sole explanation of the low-energy cutoff discussed in~\citet{PierreAuger2024jcap}.

This difference is physically reasonable. In our simulations, the fiducial magnetic field strength is in the sub-nG range and this is motivated by present constraints on the magnetisation of the cosmic web, whereas the scenarios explored by \citet{PierreAuger2024jcap} require substantially stronger local fields in the region between the observer and the nearest sources. Our results therefore support the view that the MH is likely to contribute to shaping the observed UHECR spectrum, but may not be sufficient on its own to explain the full suppression feature if the local extragalactic field remains at the level expected for realistic large-scale structure environments.

If this interpretation is correct, additional effects may need to be invoked together with the MH. These may include, for example, source-related effects, a distribution of nearby source distances, composition-dependent propagation, or self-induced confinement of escaping cosmic rays in the environments surrounding the sources~\citep{Cermenati2026aa}. 
However, a more detailed assessment of the relative importance of these effects will require dedicated simulations of the Local Universe, realistic observer locations, and an extension of the present analysis to heavier nuclei.

\section{Conclusions}
\label{sec:conclusions}

Cosmic magnetic fields play a central role in the propagation of charged cosmic rays, as they deflect particle trajectories through the Lorentz force and can reduce the maximum distance that particles are able to travel from their sources. In the ultra-high-energy regime, this effect can generate a magnetic horizon (MH), which suppresses the contribution of distant sources at sufficiently low rigidity and may therefore modify the observed UHECR spectrum and composition.

This possibility is particularly relevant in light of recent interpretations of Pierre Auger Observatory data, which indicate that unmagnetised propagation models often require very hard source spectra. In this context, the MH has been proposed as a physically motivated propagation effect that can suppress the low-rigidity flux and thereby help reconcile the data with softer source injection spectra.

In this work, we have investigated this effect using a realistic model of extragalactic magnetic fields based on state-of-the-art cosmological MHD simulations, adopting a magnetic-field configuration consistent with recent radio constraints on the magnetisation of the cosmic web. Our baseline model includes both a large-scale component, described by a primordial power spectrum $P_B(k)\propto k^{-1}$, and a small-scale contribution associated with feedback-driven magnetisation around galaxies, whose importance increases towards low redshift.

By propagating UHECR protons with the \texttt{UMAREL} code through a sequence of evolving cosmological snapshots, we find that realistic extragalactic magnetic fields produce a significant magnetic horizon below $E \lesssim 3\times10^{19}\,\rm eV$. In our fiducial model, the characteristic horizon is $R_{\rm MH}\sim 50\,\rm Mpc$ for protons arriving at $10^{18}\,\rm eV$, and $R_{\rm MH}\sim 150\,\rm Mpc$ for protons arriving at $10^{19}\,\rm eV$. At the highest energies, instead, the horizon becomes increasingly controlled by energy losses, and the difference between magnetised and unmagnetised propagation is strongly reduced.

We also showed that the topology of the magnetic field is an important ingredient. Even when the rms magnetic-field strength is kept fixed, idealised stochastic models can overestimate or underestimate the magnetic horizon by a factor of $\sim 1.5$--$2$ relative to the more complex magnetised cosmic-web distribution produced by cosmological MHD simulations. This highlights that realistic large-scale structure and magnetic-field intermittency must be taken into account in order to obtain reliable estimates of UHECR propagation effects.

Overall, our results show that observationally motivated, volume-filling extragalactic magnetic fields can produce a non-negligible magnetic horizon and should therefore be regarded as an essential ingredient in future studies of the origin and propagation of UHECRs. At the same time, our comparison with previous work suggests that the magnetic horizon alone may not fully account for the observed low-rigidity suppression, and that additional effects, such as source-environment transport or self-induced confinement, may also contribute. Extending this analysis to heavier nuclei, realistic observer locations, and constrained simulations of the Local Universe will be the next key step towards a fully consistent interpretation of UHECR data.

\begin{acknowledgements}
FV acknowledges funding under the European Union’s  Horizon Europe program through the ERC Synergy Grant COSMOMAG (Project Id. 101224803), and from  Fondazione Cariplo and Fondazione CDP, through grant n$^\circ$ Rif: 2022-2088 CUP J33C22004310003 for the ``BREAKTHRU'' project.
AF and FV acknowledge the CINECA awards ``IscrB{\_}CREW" and ``IscrC{\_}UMAREL" under the ISCRA initiative, for the availability of high-performance computing resources and support, and the usage of online storage tools kindly provided by the INAF Astronomical Archive (IA2) initiative (http://www.ia2.inaf.it). 
\end{acknowledgements}

\bibliographystyle{aa}
\bibliography{references}

\appendix

\section{Tests on UMAREL}
\label{A1}

This section is devoted to show the various tests needed to prove the physical validity of \texttt{UMAREL} and for simplicity they were conducted considering a single epoch of injection at $z\approx2$ for $\sim10^5$ protons. In particular, we binned the final evolved population of protons into four energy bins, based on their initial energy  ($10^{17},\,10^{18},\,10^{19},\,10^{20}$ eV) and, in order to monitor how the environmental conditions impact on the propagation of UHECRs, we examined the maximum distance covered by those protons in three different scenarios and compared one with each other.
Firstly, we reconstructed the fiducial setup (with $\langle B^2\rangle ^{0.5}=0.37$ nG), corresponding to a more likely view, in which protons, affected by energy losses, propagate in a realistic magnetic field. Then, we also considered two other archetypal scenarios: in the former the particles propagate in a negligible magnetic field set to $10^{-20}$ G, suffering the energy losses from the interaction with low energy photons, and in the latter, the propagation takes place in a magnetised medium where the energy losses are silenced.

\subsection{Testing the impact of the magnetic field}

\begin{figure}[ht]
    \centering
    \includegraphics[width=1\linewidth]{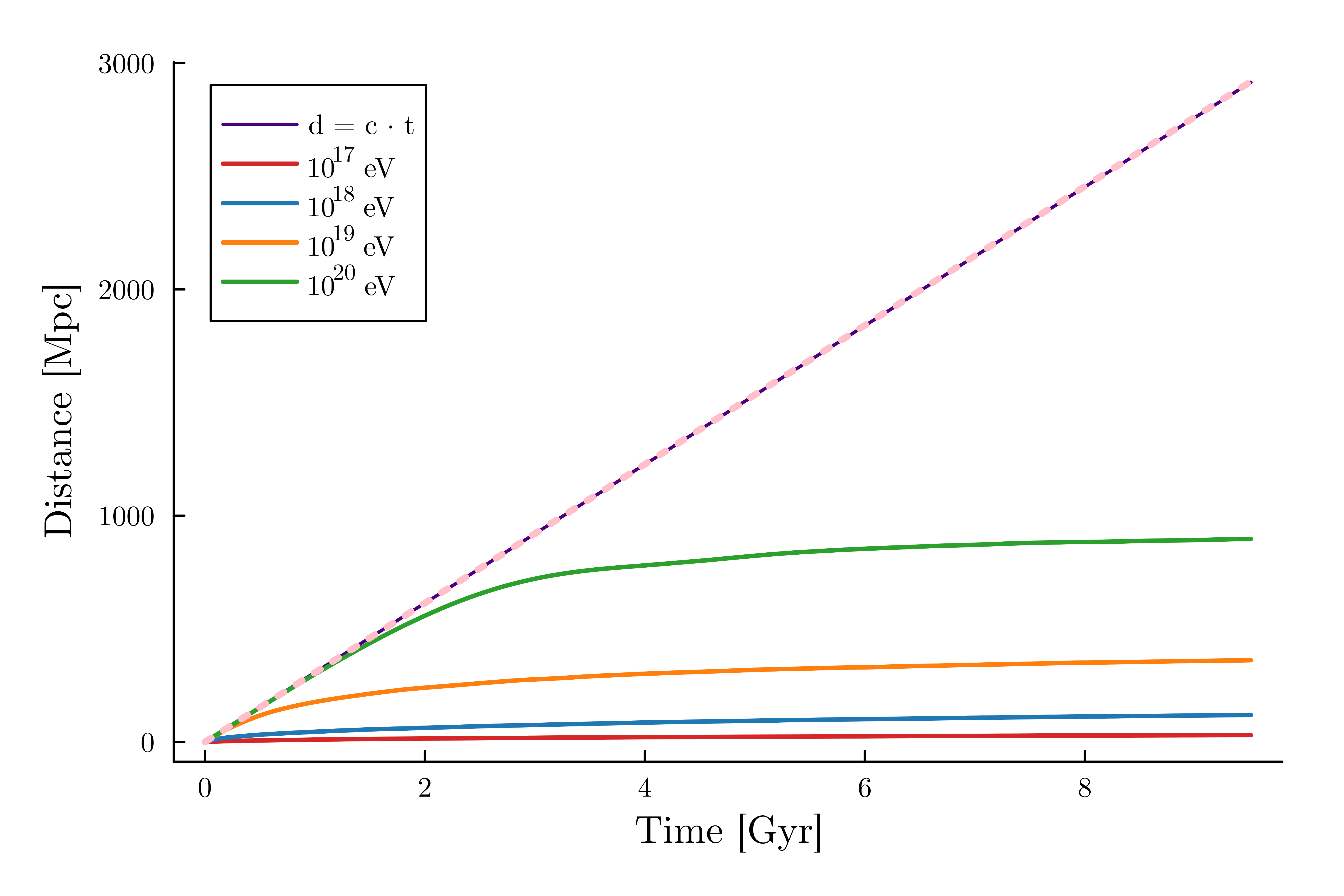}
    \caption{Comparison between the ballistic regime (dotted pink line), where a negligible magnetic field is considered, overimposed to the pink one representing the theoretical distance $d=c\cdot t$, with the fiducial case, represented by the four curves red, blue, orange and green sorted depending on the injection energy of the particle $10^{17},\,10^{18}\,,\,10^{19}$ and $10^{20}$ eV . Here the energy loss effect is taken into account. }
    \label{fig:fieldvsnofield}
\end{figure}

The first analysis is to test the impact of the magnetic field: to this end, in this section we are going to compare the realistic scenario, in which protons propagate in the magnetized medium, presented in Sec. \ref{sec:enzo} with a model where the field is set to a negligible value of $B=10^{-20}$ G. In both cases, the energy loss mechanisms due to the adiabatic expansion, pair and pion production, described in Sec.~\ref{umarel}, are considered.
Figure \ref{fig:fieldvsnofield} shows the distance travelled by the protons depending on their injection energy during 10 Gyrs of evolution, juxtaposing the fiducial case with the non magnetised one. Here, the effect of the presence of the magnetic field is tangible as in this plot the distance covered in a non magnetised environment (the dotted pink line) is compatible to a ballistic regime where the theoretical distance is $d=c\cdot t$ (purple solid line). On the other hand, when considering the propagation in a realistic magnetic field, not only is the distance at the end of the simulation significantly reduced, but it is also influenced by the energy of the particle. In fact, protons injected with a higher energy, such as $10^{20}$ eV, are able to resemble the ballistic behaviour for a longer time owing to their higher rigidity $R\propto E/Z$, whereas low energy particles, having a lower resistance against the magnetic field, present more tangled trajectories and therefore they are not able to reach further positions from their injection source.

\subsection{Testing the effects of the energy losses}

\begin{figure}[ht]
    \centering
    \includegraphics[width=1\linewidth]{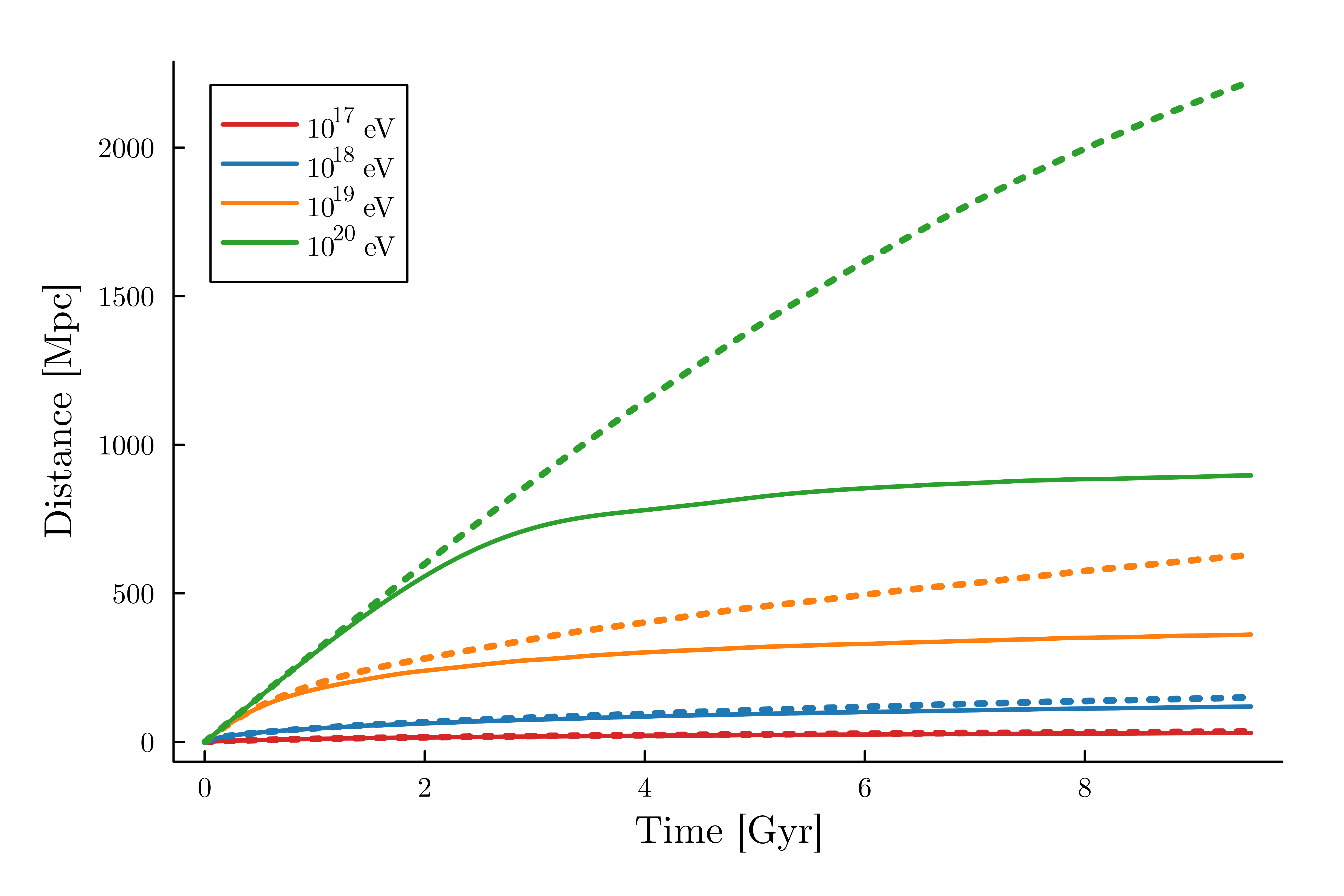}
    \caption{Comparison between the no losses regime (dotted curves), with the fiducial case (solid curves), depending on the injection energy of the particle $10^{17},\,10^{18}\,,\,10^{19}$ and $10^{20}$ eV. Here, the realistic magnetic field is present in both scenarios.}
    \label{fig:lossvsnoloss}
\end{figure}

The second test outlines the role of the energy losses on the propagation of UHECRs, that we recall being due to the adiabatic expansion and the interaction with the low energy photons of the CMB and EBL which lead to the production of pions and pairs, described in Sec.\ref{umarel}. In fact, silencing the effect of the losses mechanisms allows the cosmic protons to maintain a constant energy during their motion in the magnetic field, and so their rigidities $R\propto E/Z$ are fixed to their injection values. The comparison between the fiducial case (solid curves) with the no losses scenario (dotted curves) is depicted in Figure \ref{fig:lossvsnoloss}, where it is visible that particles able to maintain a constant energy (dotted curves) are allowed to cover larger distances. This is due to the fact that protons with a fixed energy preserve a higher resistance against the magnetic field in which they travel if compared with the ones suffering the losses and this is particularly appreciable for protons with energy of the order of $10^{20}$ eV, for which the losses are more intense, whereas it becomes less relevant as the energy of the particle decreases. 


\subsection{Testing the random injection}

\begin{figure}[ht]
    \centering
    \includegraphics[width=1\linewidth, trim = {0 0 0 1.95cm}, clip]{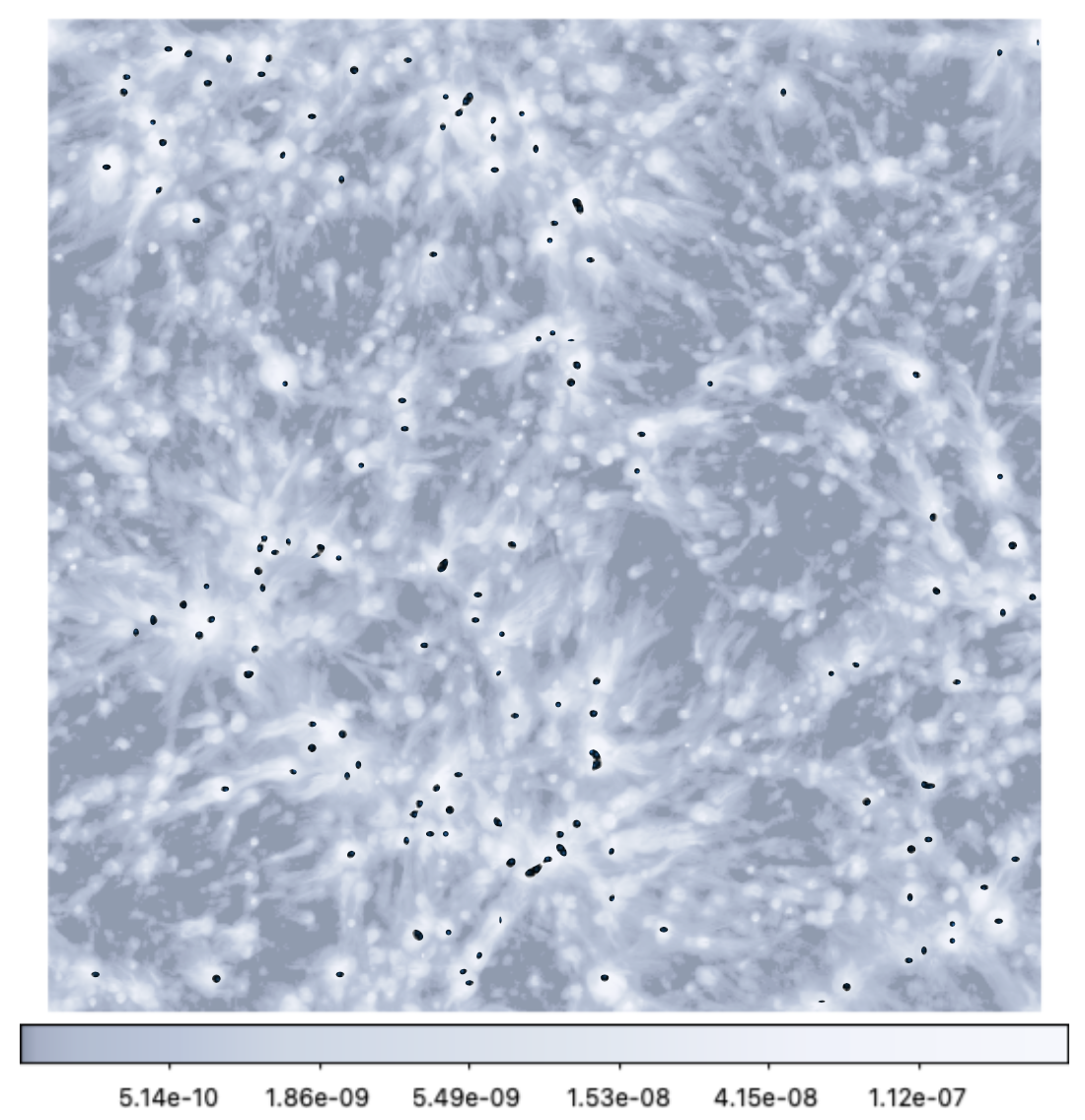}
    \caption{(Saturated) map of mean mass-weighted magnetic field strength along the line of sight (in units of comoving $\rm G$) for one snapshot of our simulation at $z \approx 0.2$, and distribution of used sources (black contours) extracted from the catalog of halos at the same redshift.} 
    \label{fig:sources}
\end{figure}

So far, the UHECRs sources has been related to all the halos with a mass larger than $10^{12}M_\odot$.
As an example, Figure \ref{fig:sources} shows the distribution of our sources of UHECRs at one of the 10 adopted injection epochs (here $z\approx0.2$), which are located within all selected halos with $M \geq 10^{12} \rm M_{\odot}$ in the simulation at this given redshift, overimposed to the projected distribution of magnetic field strength along the full volume. 

Now, for completeness we also want to investigate whether the propagation can be affected by the injection site. Thus, considering the fiducial setup, we assume that the starting position of the cosmic ray can be randomly chosen, meaning that they can start their propagation from a cosmic void as well as from a halo (and based on the volume filling factors of voids, this most often happens in the latter).
\begin{figure}[ht]
    \centering
    \includegraphics[width=1\linewidth]{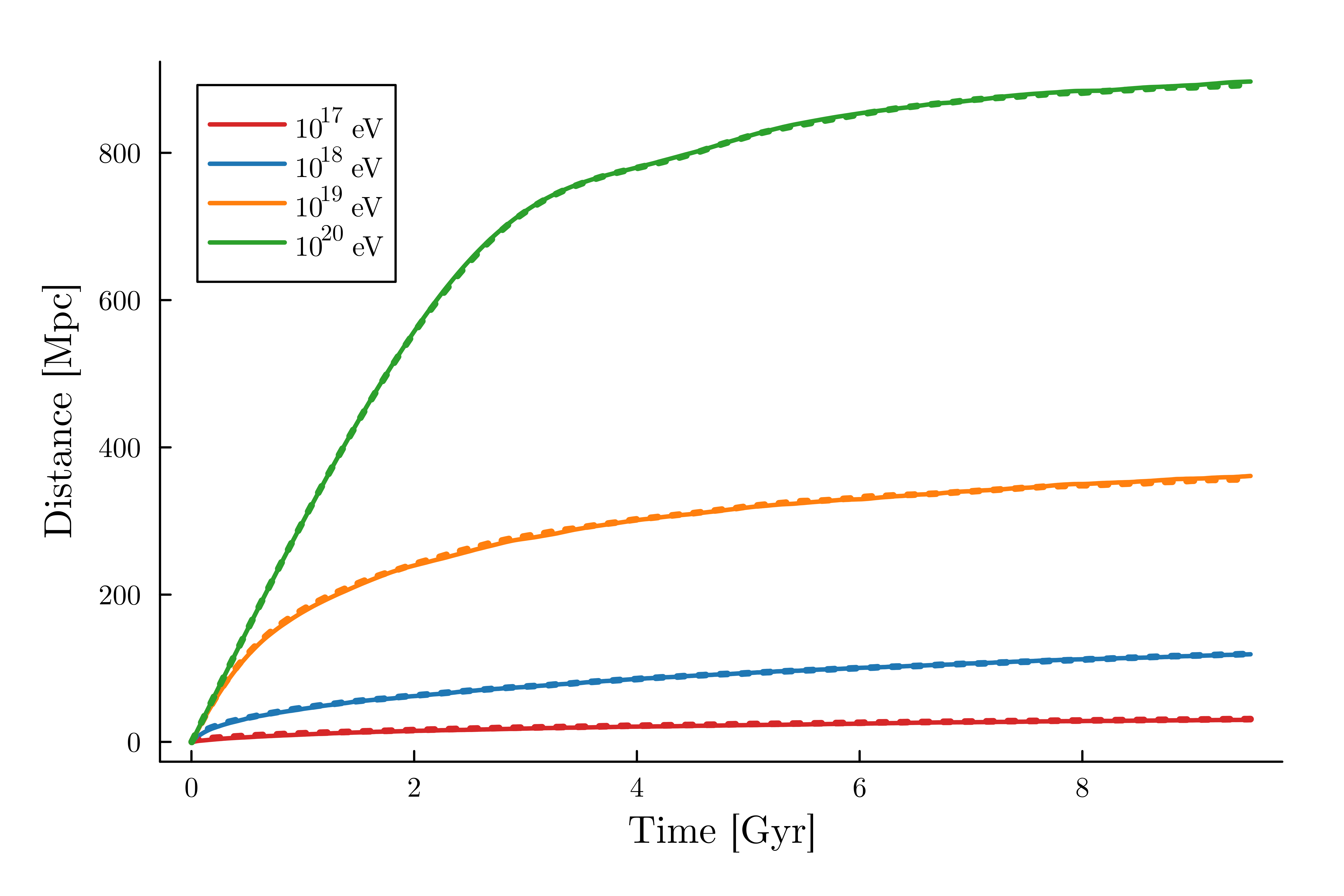}
    \caption{Comparison between the injection from haloes (solid curves)  and the random injection case (dotted curves). The distances are computed from the first time step to the last instant of the simulation for the four usual initial energy bins.}
    \label{fig:fidvsrand}
\end{figure}
We can see that the two different scenarios are identical, meaning that our estimate of the MH does not depend on the initial injection of the protons. Possible divergences, particularly visible when lowering the number of particles injected, are explained by the fact that a proton launched from a halo, thus in a highly magnetised region, is expected to reach a shorter distance than one starting from a cosmic void. 

\subsection{Test with a semi-analytically generated stochastic magnetic field} 

In this section we study more in detail the tests reported in Sec.\ref{subsec:spectra}, where we compared the propagation of UHECRs injected all once in the simulation ($z=1$) and propagated without the effect of energy losses either using the baseline magnetic field in the cosmological simulation, or a semi-analytical power-law spectrum of magnetic field with slope $P_B(k) \propto k^{-1}$ ($K-1$ model), or a semi-analytical power-law spectrum of magnetic field with slope $P_B(k) \propto k^{-11/3}$ and maximum scale $1 \rm ~Mpc$ ($K-11/3$ model). In all cases, for ease of comparison we renormalised all fields so that $\langle B^2\,\rangle^{0.5}=1$ nG in all cases.

We analysed the statistical ensemble of trajectories in all simulations, and computed the best-fit relations which best reproduce the mean squared displacement of particles from their original sources. 
We investigated for simplicity  a power-law behaviour:  $\langle r^2\rangle=D_\alpha t^\alpha$, where the two parameters $D_\alpha$ and $\alpha$, respectively the generalised diffusion coefficient and the anomalous diffusion exponent, are useful to determine different diffusive regimes. For example, the subdiffusive scheme ($\alpha<1)$ describes particles moving more slowly than in normal diffusion ($\alpha=1$) since they can be easily trapped in inhomogeneous environments. Conversely, superdiffusion ($\alpha>1$) characterises situations in which particles travel farther than in a normal diffusion process, approaching $\alpha=2$, which is typical for the ballistic regime. Within this simplistic framework, we want to test whether and how much the magnetic topology can interfere with the regime of propagations exposed above. Therefore, after having applied a fitting procedure to the actual distances, we found the two parameters $D_\alpha$ and $\alpha$, which we used to reconstruct the theoretical distances given by $y =\sqrt{D_\alpha\times t^\alpha}$. Note that, to minimise the effect of the high magnetisation that characterises the moments in which the particle is expected to be located in the proximity of its source, we have neglected the first Gyr of evolution.

\begin{figure*}[ht]
    \centering
    \includegraphics[width=.9\textwidth]{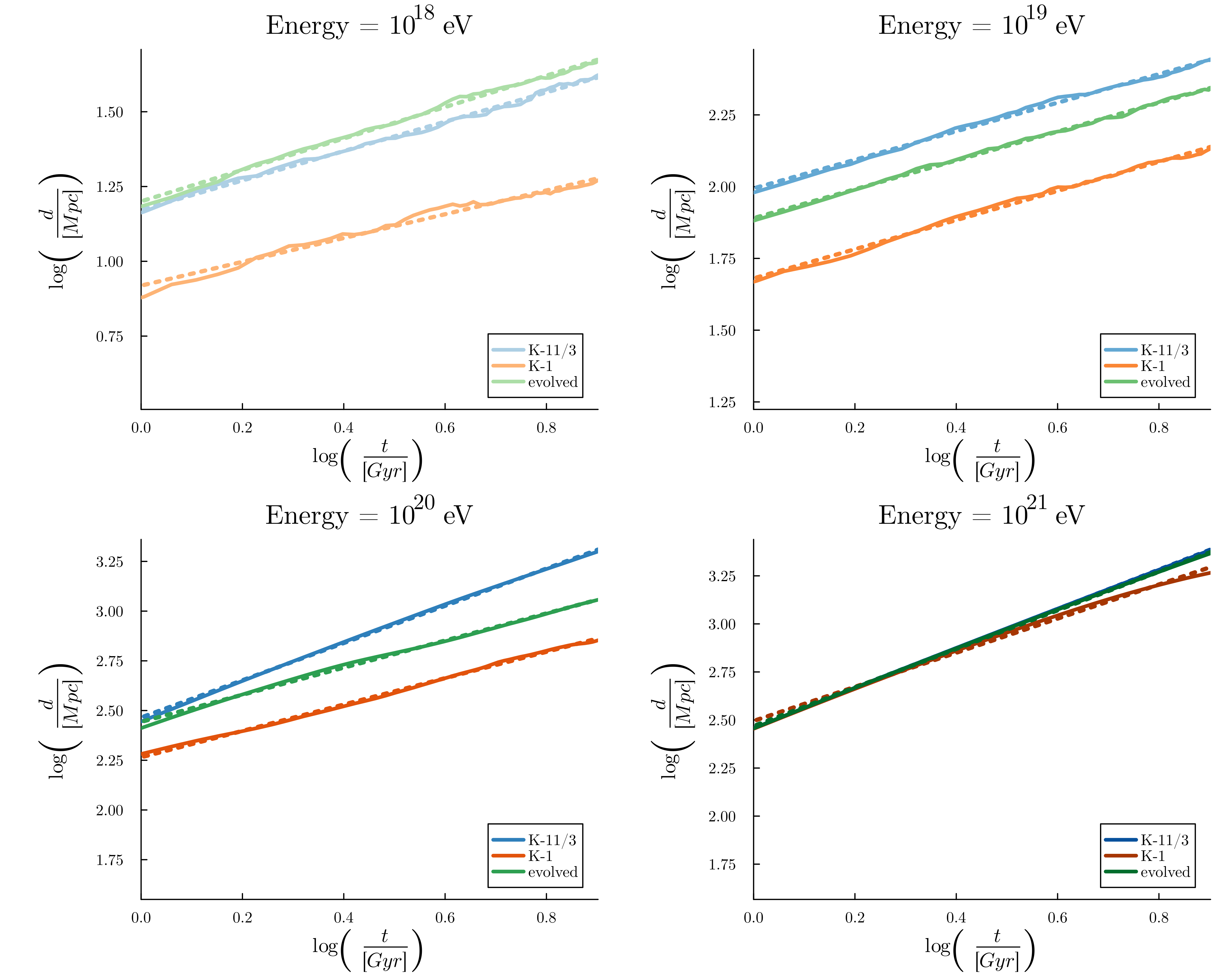}
    \caption{Comparison between the effective travelled path (solid curves) with the theoretical modelled distance (dashed curves) starting after 1 Gyrs from the injection moments neglecting the energy loss effects. Each plot concerns a different energy bin and it compares the propagation in the baseline ("evolved") magnetic field model (green coded lines), in the $P_B(k)\propto k^{-1}$  case (K-1 model, orange coded lines) and in the $P_B(k)\propto k^{-11/3}$ case with maximum coherence length 1 Mpc (K-11/3 model, blue coded lines).}
    \label{fig:fit1nG}
\end{figure*}

In Figure \ref{fig:fit1nG} we report, for each energy bin, a comparison between the theoretical distances $y =\sqrt{D_\alpha\times t^\alpha}$ (dotted curves) and the distances covered in the three different magnetised environments described above (solid curves): the evolved scenario (green shaded curves), the $K-1$ (the orange shaded curves) and the $K-11/3$ (the blue shaded curves) models.
In particular, all the distances computed here have been averaged over each population of protons injected with a certain energy ($10^{18}, 10^{19}, 10^{20}, 10^{21}$ eV) so that the significant variation depending on the magnetic setup can be appreciated; in fact particles are expected to be affected by the magnetic fluctuations with $k\sim1/r_L$, thus in a kolmogorov-like environment, the propagation is more efficient (see the blue curves at the top of each panel) since the impact of the magnetic field is less effective on small scales if compared to the $K-1$ model which represents the case where our sample of protons can reach some hundreds of Mpc at most even at the highest energy regime, as shown by the orange-theme curves in the bottom of each panel.
The distances computed in the "evolved" model (green curves) are placed in the middle of these two extremes as the magnetic field follows the dynamical evolution of structures, thus there can be regions more magnetised (such as the filaments or the sources) that limit the propagation and cosmic voids in which particles are expected to propagate farther.
Table \ref{table_fit} encodes the parameters obtained from the fitting procedure applied for each scenario, along with the scattering length scale $l=D_\alpha/c$ which represents the typical distance travelled by the proton before being deflected by the magnetic field. These results highlight the importance of accurately characterising the magnetic field structure, as the fitted parameters significantly vary depending on the chosen model, therefore, leading to different propagations regimes. 

\begin{table*}[ht]
    \centering
    \tiny
    \begin{tabular}{|c|ccc|ccc|ccc|}
    \hline Energy [eV] & \multicolumn{3}{c|}{$\alpha$} & \multicolumn{3}{c|}{$D_\alpha$ $[\mathrm{Mpc^2/Gyr}]$} & \multicolumn{3}{c|}{Length scale [Mpc] } \\
    
       & $\mathrm{K-11/3}$ & $\mathrm{K-1}$ & \textsc{evolved} & $\mathrm{K-11/3}$ & $\mathrm{K-1}$ & \textsc{evolved}&$\mathrm{K-11/3}$ & $\mathrm{K-1}$ & \textsc{evolved} \\
    \hline
        
        $10^{18}$ &$0.98$ & $0.79$ & $1.05$& $2.20\times10^2$& $6.88\times10$& $2.51\times10^2$ &$0.24$ & $7.47\times10^{-2}$&$2.73\times10^{-1}$\\
    
        \hline
        $10^{19}$ & $0.99$&  $1.01$ & $1.00$& $9.73\times10^3$ & $2.29\times10^3$ &$6.03\times10^3$& $1.05\times10$ & $2.49$&$6.55$\\
       \hline
         $10^{20}$ & $1.87$ &$1.32$& $1.36$ & $8.51\times10^{4}$ & $3.39\times10^4$& $7.69\times10^4$& $9.25\times10$ & $3.68\times10$&$8.36\times10$\\
      \hline
      $10^{21}$ & $2.04$& $1.77$&$2.00$ &$8.47\times10^{4}$ & $9.77\times10^{4}$ &$8.64\times10^{4}$& $9.21\times10$ & $1.06\times10^2$&$9.40\times10$\\
      \hline
    \end{tabular}
    \caption{Best fit values obtained from the fitting of the $y =\sqrt{D_\alpha\times t^\alpha}$ relation, starting from 1 Gyr after the injection ($\alpha$ and the length scale) with their error computed within $1\sigma$. Each parameter has been computed from the propagation in a kolmogorov-like field (K-11/3), in a synthetic field (K-1) and in dynamically evolved field (evolved), all with the same normalisation for the rms magnetic field:  $\langle B^2\rangle^{0.5}=1$ nG.}
    \label{table_fit}
\end{table*}

However, it should be stressed that this analysis represents a simplified test in which several crucial factors have been neglected (e.g. the energy loss mechanisms) and the simulation volume is relatively small ($42.5^3$ $\mathrm{Mpc^3}$), which may limit our conclusions.

\end{document}